# Whirling interlayer fields as a source of stable topological order in moiré CrI3

Doried Ghader*[1], Bilal Jabakhanji[1], and Alessandro Stroppa[2]

[1] College of Engineering and Technology, American University of the Middle East, Egaila 54200, Kuwait

[2] CNR-SPIN c/o Department of Physical and Chemical Sciences, University of L'Aquila, Via Vetoio, I-67100 Coppito, L'Aquila, Italy

*doried.ghader@aum.edu.kw

**Abstract.** The moiré engineering of two-dimensional magnets opens unprecedented opportunities to design novel magnetic states with promises for spintronic device applications. The possibility of stabilizing skyrmions in these materials without chiral spin-orbit couplings or dipolar interactions is yet to be explored. Here, we investigate the formation and control of ground state topological spin textures (TSTs) in moiré $CrI_3$ using stochastic Landau-Lifshitz-Gilbert simulations. We unveil the emergence of interlayer vortex and antivortex Heisenberg exchange fields, stabilizing spontaneous and field-assisted ground state TSTs with various topologies. The developed study accounts for the full bilayer spin dynamics, thermal fluctuations, and intrinsic spin-orbit couplings. By examining the effect of the Kitaev interaction and the next nearest-neighbor Dzyaloshinskii-Moriya interaction, we propose the latter as the unique spin-orbit coupling mechanism compatible with experiments on monolayer and twisted $CrI_3$. Our findings contribute to the current knowledge about moiré skyrmionics and uncover the nature of spin-orbit coupling in $CrI_3$.

## Introduction

Topological magnetism is an intriguing field on the frontier of condensed matter physics with great promises for future information technology[1–4]. Skyrmion[5], a particle-like spin-whirling vortex, was the first class of topological spin structures (TSTs) to be realized experimentally in the chiral magnet $MnSi$[6]. While skyrmions remain a prominent example of TSTs, several alternatives have been predicted and observed in recent years, such as antiskyrmions[7], biskyrmions[8,9], skyrmioniums[10,11], and bimerons[12,13]. Generally, TSTs are robustly stable with particle-like properties due to their topological protection. They carry a quantized topological charge and can interact via attractive and repulsive forces. The topological charge quantifies the real-space nontrivial topology of the spins, and it is defined as[14–16]

$$Q = \frac{1}{4\pi} \int d^2r [\partial_x \mathbf{S} \times \partial_y \mathbf{S}] \cdot \mathbf{S}$$

(1)

where $\mathbf{S}(\mathbf{r})$ denotes the spin density field. In Equation (1), the term $\partial_x \mathbf{S} \times \partial_y \mathbf{S}$ represents the vorticity of the spin texture and is determined by the in-plane components of the boundary spins. The vorticity classifies out-of-plane spin textures into skyrmions and antiskyrmions, with vortex and antivortex profiles, respectively[17]. Based on their whirling profile, skyrmions can be further classified as Bloch-type and Néel-type. Precisely, in a Bloch-type (Néel-type) skyrmion, the spins rotate perpendicular to (along) the radial direction when moving from the core to the periphery. Other significant properties of the TST morphology are the polarity (alignment of the core spins) and helicity (the global rotation angle around the out-of-plane axis).

Chiral interactions in magnetic films stabilize TSTs with fixed chirality (fixed vorticity and helicity)[6,17–23]. The most prominent example of chiral interactions is the NN DMI, arising in noncentrosymmetric lattices lacking space-inversion symmetry. Moreover, TSTs can form in centrosymmetric materials due to nonchiral interactions[24–35], such as dipolar interactions[24,26,30], magnetic anisotropy[29,34], quantum fluctuations[36], and static random fields[34,35]. In the nonchiral magnetic films, the vorticity and helicity act as additional degrees of freedom, relevant for spintronic and topological applications[30–33]. An external magnetic field usually assists the formation of TSTs in chiral and nonchiral magnetic films. On the other hand, materials hosting spontaneous TSTs are rare. So far, theoretical research has predicted spontaneous TSTs in itinerant magnets with high-order spin interactions[19,21,37–39] and $NiI_2$ monolayers[40] with anisotropic exchange interaction.

The search for topological magnetic order recently extended to the newly discovered two-dimensional (2D) magnets. Experimental research has reported TSTs in 2D layered magnetic materials and heterostructures[41–44]. Parallel to the experimental investigations, intensive theoretical efforts have predicted TSTs in chiral 2D magnets, including Janus monolayers[45–47], multiferroics[48–50], and monolayers with in-plane magnetic order[51,52].

Furthermore, 2D magnets offer unique opportunities to engineer TSTs via the modulated interlayer coupling in the moiré superlattices of twisted or mismatched bilayers. The mechanism was initially discussed in a heterostructure formed of a nonchiral ferromagnetic (FM) monolayer on an antiferromagnetic (AFM) substrate[53]. Bloch-type skyrmions emerge from the registry-dependent interlayer exchange and dipolar couplings in the heterostructure. Several theoretical studies followed, exploring TSTs in mismatched and twisted magnetic bilayers[54–58]. Akram *et al.*[56] and Hejazi *et al.*[57] reported skyrmions in the chiral FM-AFM heterostructure. Ray *et al.*[58] explored the TSTs in nonchiral moiré magnets with interlayer dipolar interactions. Moreover, Xiao *et al.*[55] predicted metastable skyrmions as excited states in twisted chromium trihalides $CrX_3$ bilayers ($X = I, Br$) without NN DMI or dipolar interactions. The effect of NN DMI in twisted $CrX_3$ ($X = I, Br, Cl$) was addressed in a recent work[59], reporting Néel-type skyrmions in $CrI_3$ and $CrBr_3$.

The theoretical predictions on moiré magnets are reinforced by recent experiments on twisted $CrI_3$ bilayers [60,61]. Xu *et al.*[60] reported coexisting FM-AFM states in twisted bilayer $CrI_3$ at small twist angles ($1° \lesssim \theta \lesssim 4°$). The FM (AFM) domains form at the local rhombohedral (monoclinic) stackings in the moiré supercell. However, the AFM domains are found to disappear at relatively larger twist angles ($\theta \gtrsim 4°$), resulting in a pure FM ground state. On the other hand, Song *et al.* [61]

targeted slightly twisted $CrI_3$ bilayers with $\theta \lesssim 0.5°$. Due to the lattice reconstruction at tiny twist angles, the authors observed a substantially disordered moiré pattern and a complicated magnetic structure.

From the theoretical side, the methods used so far to determine the magnetic textures in moiré magnets are the Landau-Lifshitz-Gilbert (LLG) approach[53,55,56,60], the continuum low-energy field theory[54,57], and Monte Carlo simulations with continuum interlayer exchange[58]. Nevertheless, an atomistic theoretical model that incorporates thermal fluctuations has not been reported yet. Moreover, existing studies on moiré magnets (except Akram *et al.*[59]) oversimplify the moiré bilayer as a monolayer with a spatially modulated external magnetic field. This approximation assumes fixed spins in one of the layers. Akram *et al.*[59] included the spin dynamics in both layers but with an interlayer interaction taken only between nearest neighbors. Additionally, previous studies on moiré $CrI_3$[54,55,58,59] adopt the Heisenberg model, excluding spin-orbit interactions like the intrinsic next NN DMI (IDMI) and the Kitaev interaction. However, the pure Heisenberg model is inadequate for $CrI_3$ since it fails to explain the gapped magnon spectrum[62–64] in this material. Meanwhile, both the Heisenberg-IDMI[63,64] and Heisenberg-Kitaev[62,63] models can reproduce the spectrum, creating wide controversy on the true microscopic spin Hamiltonian for $CrI_3$[65]. Hence, it is important to investigate the Heisenberg-IDMI and Heisenberg-Kitaev models for moiré $CrI_3$ and check if they can reproduce the recent experimental observations[60].

In this work, we predict the emergence of whirling interlayer exchange fields in twisted moiré $CrI_3$. At long moiré periodicity, the interlayer interaction dominates, and the spins align with the nontrivial moiré fields, inducing stable TSTs. Hence, the whirling moiré fields uncover a mechanism to stabilize ground state TSTs without the need for NN DMI or dipolar interactions.

The presented study mimics realistic experiments by cooling the bilayer system from initial paramagnetic states. We use the stochastic Landau-Lifshitz-Gilbert (sLLG) equation to account for thermal effects. We include both layers in the spin dynamics simulations after developing an atomistic approach that accounts for the effective interlayer coupling beyond the NN approximation. Further, the study covers and compares the chiral and nonchiral Heisenberg, Heisenberg-IDMI, and Heisenberg-Kitaev models.

In the absence of the chiral NN DMI, the vorticity and helicity of the interlayer fields act as degrees of freedom. As a result, the stochastic time evolution of the interlayer fields does not follow deterministic rules in the nonchiral models. Instead, the interlayer fields' profiles at low temperatures depend crucially on the initial paramagnetic state. As a result, nontrivial moiré fields with various chiralities can emerge at small twist angles to stabilize a zoo of TSTs. On the contrary, a sizeable chiral NN DMI locks the moiré interlayer field's chirality, resulting in Néel-type skyrmions for any initial paramagnetic state.

The study covers the field-assisted and spontaneous formation of the TSTs. Cooling with an external magnetic field traps the TSTs at the monoclinic AFM regions of the moiré supercell. Conversely, spontaneous TSTs can emerge in the FM regions, AFM regions, or a combination of FM and AFM regions. Moreover, spontaneous spin textures can merge to form magnetic strips with chiral domain walls. Both the spontaneous and the field-assisted TSTs can be drastically manipulated by a new magnetic field applied at $0\,K$.

Finally, we show that moiré engineering provides insights into the fundamental interactions underlying 2D magnets. Specifically, we consider the recent experimental results[60] on moiré $CrI_3$ to address the controversy regarding the correct microscopic model for $CrI_3$. We explore the twist and temperature ($T$) dependence of the averaged magnetization ($M$) in the Heisenberg-IDMI and Heisenberg-Kitaev models. The Heisenberg-IDMI model is found to display a twist-dependent $M - T$ curve that evolves towards an FM ground state at large angles ($\theta \gtrsim 4.3°$), in agreement with the experimental results[60]. However, the Heisenberg-Kitaev model fails to describe the twist-dependent ground state. Therefore, we conclude that the Heisenberg-IDMI model is the unique model that can reproduce the experimental results in monolayer and moiré $CrI_3$.

## Results

### Spin Hamiltonians

We adopt a generic Hamiltonian for twisted $CrI_3$ bilayers including exchange, DM, and Kitaev interactions,

$$\mathcal{H} = -J \sum_{l,i,j} \mathbf{S}_{li}.\mathbf{S}_{lj} - \mathcal{A} \sum_{l,i} (\mathbf{S}_{li}.\hat{\mathbf{z}})^2 + D \sum_{l,m,n} \hat{\mathbf{D}}_{mn}.\mathbf{S}_{lm} \times \mathbf{S}_{ln} + d \sum_{l,m,n} \hat{\mathbf{d}}_{ij}.\mathbf{S}_{li} \times \mathbf{S}_{lj}$$
$$- \sum_{i,j} J_\perp(\mathbf{r}_{ij}) \mathbf{S}_{2i}.\mathbf{S}_{1j} - \sum_{l,i} \mathbf{B}.\mathbf{S}_{li} - \sum_l \sum_{\langle i,j \rangle \in \lambda\mu(\nu)} \left[ K S_i^\nu S_j^\nu + \Gamma\left( S_i^\lambda S_j^\mu + S_i^\mu S_j^\lambda \right) \right]$$

(2)

The vector $\mathbf{S}_{li}$ denotes the classical spin on a site $i$ in layer $l$, with position vector $\mathbf{r}_{li}$. We set $l = 1, 2$ for the bottom and top layers, respectively. $J$ and $\mathcal{A}$ are the NN intralayer Heisenberg coupling and the easy axis magnetic anisotropy, respectively. The third term in $\mathcal{H}$ is the nonchiral Néel-type IDMI[63,64,66–69] between next NN illustrated in Supplementary Figure 1a. In contrast, the fourth term accounts for possible NN chiral DMI (Supplementary Figure 1b) induced by the broken inversion symmetry in the twisted system. $J_\perp(\mathbf{r}_{ij}) = J_\perp(\mathbf{r}_{2i} - \mathbf{r}_{1j})$ is the distance-dependent interlayer coupling between spins $\mathbf{S}_{2i}$ and $\mathbf{S}_{1i}$. The sixth term is the Zeeman coupling by an external magnetic field $\mathbf{B}$ normal to the bilayer. The last term is the Kitaev Hamiltonian[62,69–71], parametrized by $K$ and $\Gamma$. All terms in $\mathcal{H}$ are expressed in the coordinate axes illustrated in Fig.1a, except the Kitaev Hamiltonian, which is expressed in the octahedral coordinate axes[62,63]. Details on the octahedral coordinate axes can be found, for example, in Chen *et al.*[63]. The triplet $(\lambda, \mu, \nu)$ in the summation represents any permutation of the octahedral coordinates.

Table 1 reports the proper parametrization of $\mathcal{H}$ to reproduce the Heisenberg, Heisenberg-IDMI, and Heisenberg-Kitaev models. Further, we will denote models with sizeable (respectively negligible) NN DMI as chiral (nonchiral).

**The atomistic interlayer coupling approach**

$CrI_3$ is the most prominent example of stacking-dependent magnetism in 2D magnets and constitutes an excellent candidate to discover the intriguing physics underlying moiré magnets. In the twisted bilayer, the $CrI_3$ moiré superlattice hosts regions with local rhombohedral (AB, BA, and AA) and monoclinic ($\mathcal{M}$) stackings (Figure 1a). The interlayer exchange is FM (AFM) at the rhombohedral (monoclinic) stackings[72–76]. Several authors calculated the stacking-dependent interlayer exchange energy in $CrI_3$[55,59,74]. Here, we adopt the DFT results obtained by Xiao *et al.*[55] and develop a method to determine the atomistic interlayer exchange coupling accordingly. This approach will later allow us to simulate the time evolution of the interlayer exchange fields.

Figure 1b presents the moiré interlayer exchange energy $E_{int}$ (Xiao *et al.*[55]), characterized by three AFM monoclinic regions labeled I, II, and III. Without loss of generality, we choose a spin $\mathbf{S}_{2i}$ at position $\mathbf{r}_{2i}$ in layer 2 (top layer). The spatially modulated effective interlayer coupling can be expressed as[55] $J_\perp^{eff}(\mathbf{r}_{2i}) = \sum_j J_\perp(\mathbf{r}_{2i} - \mathbf{r}_{1j}) = -E_{int}(\mathbf{r}_{2i})/4S^2$. Here, $S = 3/2$ is the spin of the $Cr$ magnetic atom. Next, we assume $J_\perp(\mathbf{r}_{2i} - \mathbf{r}_{1j})$ decays exponentially[56,57] as a function of the distance, that is

$$J_\perp(\mathbf{r}_{2i} - \mathbf{r}_{1j}) = -\frac{E_{int}(\mathbf{r}_{2i})}{4S^2} e^{-\delta_{2i}\sqrt{|\mathbf{r}_{2i}-\mathbf{r}_{1j}|+r_0^2}}$$

(3)

where $r_0$ denotes the interlayer separation. Then, the decay factor $\delta_{2i}$ (and consequently $J_\perp(\mathbf{r}_{2i} - \mathbf{r}_{1j})$) can be determined numerically for every spin $\mathbf{S}_{2i}$ in the moiré supercell by solving the equation,

$$\sum_j e^{-\delta_{2i}\sqrt{|\mathbf{r}_{2i}-\mathbf{r}_{1j}|+d^2}} = 1$$

(4)

In particular, we solve Equation (4) for a large cutoff interlayer interaction radius $|\mathbf{r}_{2i} - \mathbf{r}_{1j}| \leq a$ ($a$ is the $CrI_3$ lattice constant) to ensure adequate distribution of the effective interlayer interaction

over the interlayer sites, while the exponential term cancels irrelevant contributions. This is desired to avoid biased interlayer fields in the atomistic sLLG simulation.

The numerical approach eventually yields the distance-dependent interlayer coupling $J_\perp$ for all interacting spins in layers 1 and 2. The effective moiré interlayer field on a specific site $i$ in layer 2 can be deduced as $\sum_j J_\perp(\mathbf{r}_{2i} - \mathbf{r}_{1j}) \mathbf{S}_{1j}$. Similarly, for a specific site $j$ in layer 1, the moiré field is $\sum_i J_\perp(\mathbf{r}_{2i} - \mathbf{r}_{1j}) \mathbf{S}_{2i}$.

Finally, we note that Equation 3 assumes an isotropic form for $J_\perp(\mathbf{r}_{2i} - \mathbf{r}_{1j})$. Further improvement to include anisotropic terms such as an interlayer spin-orbit coupling might be interesting, requiring future DFT investigations.

**The stochastic LLG approach**

We used the Vampire atomistic spin dynamics software[77] to determine the time evolution of the spin textures using the sLLG equation. Generally, atomistic simulations better account for the complex stacking-dependent magnetism in moiré $CrI_3$ compared to the continuum approach. The input files for the Vampire simulations were prepared using the Mathematica software[78].

The sLLG equation reads

$$\frac{\partial \mathbf{S}_{li}}{\partial t} = -\frac{\gamma}{1+\lambda^2}\left[\mathbf{S}_{li} \times \mathbf{H}_{li}^{eff} + \lambda \mathbf{S}_{li} \times \left(\mathbf{S}_{li} \times \mathbf{H}_{li}^{eff}\right)\right]$$

(5)

where $\gamma$ and $\lambda$ are the gyromagnetic ratio and Gilbert damping, respectively. $\mathbf{H}_{li}^{eff}$ is the net magnetic field on spin $\mathbf{S}_{li}$,

$$\mathbf{H}_{li}^{eff} = -\frac{1}{\mu_s}\frac{\partial \mathcal{H}}{\partial \mathbf{S}_{li}} + \mathbf{H}_{li}^{th}$$

(6)

The first term in $\mathbf{H}_{li}^{eff}$ can be derived from Equations (2), (3), and (4), while $\mathbf{H}_{li}^{th}$ is the effective thermal field included in Vampire using the Langevin Dynamics method[79].

To mimic real experiments, we launch the sLLG simulations with random initial spins at a high temperature ($50\ K$), followed by gradual cooling to $0\ K$. The system is cooled with $3 \times 10^7$ total time steps. We use $1 \times 10^{-16} s$ for the time step and a cooling time of $1 ns$. The spin configurations are determined in both layers at different temperatures, down to the ground state ($0\ K$), using the Heun integration scheme[79] and imposing periodic boundary conditions. The Heun integration scheme is preferred for stochastic spin dynamics due to its computational efficiency and accuracy in reproducing the correct ground state[77].

We analyzed the ground state spin textures at commensurate twist angles in the range $0.65° \leq \theta \leq 6°$, which is relevant to the recent experimental work[60]. We will focus more on the Heisenberg and the Heisenberg-IDMI models since the Heisenberg-Kitaev model was found inconsistent with the experimental results on moiré $CrI_3$.

**Cooling with an applied magnetic field**

We start the discussion by considering the field-assisted TSTs in the nonchiral Heisenberg and Heisenberg-IDMI models. For slight twists, the interlayer interaction dominates the intralayer exchange[54,60] and induces three magnetic bubbles in the monoclinic AFM regions of the moiré. Consequently, simulating the time evolution of the interlayer exchange fields is crucial to investigate the possible emergence of stable TSTs.

The monolayer approximation freezes $\mathbf{S}_{1j}$ along $+\hat{\mathbf{z}}$ and yields a collinear interlayer field on the top layer, aligned along $\pm\hat{\mathbf{z}}$. Consequently, the trivial interlayer field in the monolayer approximation does not favor the formation of stable TSTs in the absence of the NN DMI[55]. Here, we reveal a different picture where the moiré interlayer fields acquire nontrivial profiles, stabilizing ground state TSTs.

In our approach, the orientation of the interlayer field on $\mathbf{S}_{2i}$, $\sum_j J_\perp(\mathbf{r}_{2i} - \mathbf{r}_{1j})\mathbf{S}_{1j}$, is determined by the sign of $J_\perp(\mathbf{r}_{2i} - \mathbf{r}_{1j})$ and the directions of the contributing spins $\mathbf{S}_{1j}$. Similarly, for the interlayer field on layer 1. Consequently, in the stochastic description of the moiré spin dynamics, the interlayer fields act as dynamic (time-dependent) random fields during the cooling process before reaching static configurations near $0\ K$ (Supplementary Movie 1). The thermal fluctuations dominate the magnetic interactions down to the ordering temperature and play an important role in the time evolution of the moiré interlayer fields. However, the thermal fluctuations diminish at low temperatures, and the competing magnetic interactions gradually dominate below the ordering temperature. The combined effects of the thermal fluctuations and the competing magnetic interactions shape the moiré field during the cooling process, and the moiré fields can converge to nontrivial textures at $0\ K$.

Figure 1c illustrates an example of antivortex interlayer fields emerging in the AFM regions of the moiré. The figure simulates the moire field on the bottom layer of the nonchiral Heisenberg model with $\theta = 1.35°$ and $B = 1\ T$. Since the interlayer interaction dominates the intralayer exchange at long moiré periodicity, the spins align with the interlayer field to minimize the energy. Consequently, the interlayer field shapes the spin textures' morphology in the AFM regions. Specifically, the spins inherit the vorticity and helicity of the antivortex interlayer field (Figure 1c), stabilizing three ground state antiskyrmions (Fig. 1d).

In the nonchiral Heisenberg and Heisenberg-IDMI models, the vorticity and helicity are degrees of freedom. The time evolution of the interlayer fields does not follow deterministic rules that can predict the stochastic simulation's results a priori. The profile of the final interlayer fields and the corresponding magnetic ground state depend crucially on the initial paramagnetic state, chosen randomly in our simulations. Any combination of the degrees of freedom is allowed and can be probed by different initial states. Accordingly, topological and trivial MBs can coexist in the moiré bilayer (Supplementary Figure 2a, b), with arbitrary distribution over the layers. For completeness, examples of trivial interlayer fields stabilizing non-topological MBs are presented in Supplementary Figures 2c and 2d. Generally, the spins interacting with a topological or trivial MB in the opposite layer are twisted relative to $+\hat{\mathbf{z}}$.

The crucial dependence of the ground state on the initial paramagnetic configuration is elaborated in Supplementary Figure 3. Therefore, a reliable study requires exhaustive numerical experiments, including several initial random spin configurations. We studied bilayers with commensurate angles $0.65° \leq \theta \leq 6°$ and external magnetic fields in the range $100\ mT \leq B \leq 1.5\ T$, varied in steps of $100\ mT$. We additionally included the magnetic fields $0.25\ T$, $0.75\ T$, and $1.25\ T$. We tested six distinct random initial states for each twist angle and magnetic field, generating 108 different simulations for a given twist angle. Similar to previous theoretical works[55,58,59], we present results for simulations over a single moiré supercell. Nevertheless, we have extensively investigated samples with multiple moiré supercells. We found that including additional moiré supercells in the simulation of field-assisted TSTs is equivalent to changing the initial random spin configurations on a single moiré supercell. Meanwhile, we stress that the moiré-periodicity of the magnetic ground state is broken in multi-moiré supercell samples. In particular, since adjacent moiré supercells have distinct initial random spin configurations, they converge to ground states with different types of MBs (topological or trivial) in the AFM regions.

The low-temperature interlayer fields manifest in various profiles, inducing multi-flavored TSTs, such as antiskyrmions (Fig. 1c, d), Néel-type skyrmions (Fig. 2a, b), Bloch-type skyrmions (Fig. 2c, d), and topological defects with $|Q| > 1$ (Fig. 2e, f). The TSTs are observed in the range $\theta \lesssim 2.13°$ as a rough estimation. The high topological charge ($Q = 2$) spin texture in Figure 2e can be interpreted as a magnetically stable bound state of two antiskyrmions with opposite helicities. Bound states of two skyrmions with opposite helicities ($Q = -2$) are also possible in the moiré bilayer, and an example is presented in Supplementary Figure 4e. The formation of such skyrmionic molecules stems from the helicity degree of freedom and can be realized only in nonchiral magnets. They are analogous to bi-skyrmions and bi-antiskyrmions observed in nonchiral frustrated magnetic films[30–33], but with zero separation between the antiskyrmions[33].

The rich spectrum of TSTs in the nonchiral Heisenberg and Heisenberg-IDMI models is further elaborated in Table 2. In particular, we present the ground state with the maximum number of TSTs from six trials performed for each angle and magnetic field. For example, we choose to present the result of simulation 1 from Supplementary Figure 3 because it displays three TSTs for the Heisenberg model with $\theta = 1.35°$ and $B = 0.25\ T$. This criterion is occasionally dropped in Table 2 to report TSTs with a high topological charge. Therefore, the results of Table 2 are to be interpreted as insightful rather than deterministic results.

It can be noticed from Table 2 that TSTs can be realized even at relatively large angles (e.g., $\theta = 2.13°$) by varying the magnetic field. Moreover, simulations based on the nonchiral Heisenberg and Heisenberg-IDMI models yield different results for the same initial random state, indicating that the IDMI affects the spin dynamics in the bilayer system. Nevertheless, the overall results and topological spectra are comparable for the nonchiral Heisenberg and Heisenberg-IDMI models, suggesting that the IDMI interaction does not have characteristic signatures on the morphology of the TSTs.

We verified the robustness of the TSTs when the external magnetic field is turned off at $0\ K$ and observed stability in the TSTs' morphology with no effect on the topological charge (Supplementary Figure 4 a-d). Consequently, the textured interlayer interaction stabilizes the topological order without a permanent external field, which is desired for skyrmion-based spintronic devices. In experiments, adjacent moiré supercells will have distinct initial states and converge to different ground states. Nevertheless, our results suggest that successive heating-cooling trials and magnetic field manipulation can experimentally establish a topologically rich ground state, with skyrmions, antiskyrmions, high topological charge spin textures, and a minimal number of trivial magnetic bubbles. Such ground states with coexisting distinct TSTs are challenging to realize in conventional materials and are in-demand for memory and logic applications[80].

Unsurprisingly, a sizeable NN DMI locks the chirality of the TSTs in the Heisenberg and the Heisenberg-IDMI models, hence producing deterministic results. Specifically, the chiral models present Néel-type skyrmions with a fixed topological charge $Q = -1$ for any initial paramagnetic spin configuration (Supplementary Figure 1c, d). The Néel-type skyrmions can form in any layer, and the moiré-periodicity is broken in multi-moiré supercell samples. Our calculations assumed a sizeable NN DMI $d = 0.15\ meV$, stabilizing Néel-type skyrmions in the range $\theta \lesssim 3.15°$.

The MBs in the chiral/nonchiral Heisenberg and Heisenberg-IDMI models are found to disappear at relatively large angles ($\theta \gtrsim 4.2°$), and the bilayer converges systematically towards an FM state as we increase the twist angle. Further discussions are presented in the last part of the Results section.

We proceed to discuss the possibility of manipulating the topological ground state at $0\ K$. Applying a new magnetic field in the direction of the MBs' core spins promotes a controllable outward motion of the domain walls without affecting the topological charge. Consequently, it is possible to inflate and couple the spin textures initially trapped in the AFM regions to realize skyrmion pairs (Supplementary Figure 5a, b), antiskyrmion pairs (Supplementary Figure 5d, e), and TST-

(trivial) MB pairs (Supplementary Figure 5g, h). Moreover, a sufficiently large magnetic field inflates the MBs drastically and eventually induces a global magnetization reversal (Supplementary Figure 5c, f, i and Supplementary Movie 2). As a result, a final topological ground state is achieved with reversed spins compared to the initial state. Generally, the MBs jump to the opposite layer in the final ground state, and the spin reversal can modify the TST's vorticity and helicity (Supplementary Figure 5). We note the related discussion in Xiao *et al.*[55], based on the monolayer approximation that cannot capture the complete picture presented here. In particular, the magnetization reversal erases the MBs in the monolayer approach since they cannot form in the opposite layer (assumed with fixed spins).

The stabilization mechanism for TSTs via whirling moiré interlayer fields is general and applies to the nonchiral Heisenberg-Kitaev model. The TSTs' charges in the Heisenberg-Kitaev model are comparable to the previous models (Table 2). However, unlike the IDMI, the Kitaev interaction has clear signatures on the ground state TSTs' chirality. The Kitaev interaction induces a canting-like effect, where neighboring peripheral spins can cross to form pairs of frustrated spins (Supplementary Figure 6 a-d). Therefore, the nonchiral Heisenberg-Kitaev model displays TSTs with unconventional morphologies compared to the Heisenberg and the Heisenberg-IDMI models. The canting-like effect persists in the chiral Heisenberg-Kitaev model, and ideal Néel-type skyrmions were not observed even for a sizeable NN DMI (Supplementary Figure 6e, f).

The Heisenberg-Kitaev model does not transfer to an FM bilayer, and the MBs survive at large angles. The in-plane Heisenberg interaction is weak (Table 1) and is not expected to dominate the local AFM coupling even at large angles. Therefore, the observed behavior suggests that the interlayer AFM interaction is dominant over the Kitaev in-plane interaction throughout the range $0.65° \leq \theta \leq 6°$.

**Cooling without an applied field**

Whirling interlayer fields can also stabilize spontaneous TSTs without an external magnetic field in the three models. Unlike the field-assisted TSTs, which are confined to the AFM regions, the spontaneous TSTs can form in any region of the moiré supercell. In particular, the spontaneous TSTs can emerge in the AFM regions, FM regions, or a combination of the AFM and FM regions, depending on the initial spin configuration (Fig. 3). The merging of spin textures over the AFM and FM regions can generate giant spontaneous TSTs at slight twists (Supplementary Figure 7). Moreover, the spin textures can merge across the entire moiré supercell (Fig. 3) to form magnetic strips in one or both layers. These wavy strip-shaped magnetic domains are separated by chiral domain walls and can develop in different directions depending on the particularly merged MBs. Figure 3 illustrates the crucial dependence of the spontaneous spin textures on the initial random configuration. Therefore, similar to the discussion in the previous section, the moiré-periodicity of the spontaneous spin textures is broken since adjacent moiré supercells have different initial states.

Nevertheless, the single-moiré supercell simulations remain faithful to reproduce the main features of the spontaneous spin textures. Our intensive numerical investigation of multi-moiré supercell samples did not disclose substantially new information, except that the moiré-periodicity is broken in such samples.

In the chiral/nonchiral Heisenberg and Heisenberg-IDMI models, the various merging scenarios are promoted at small angles ($\theta \lesssim 2.44°$), whereas the spin textures are confined to the AFM regions at larger angles. Moreover, our previous discussion regarding the degrees of freedom in the nonchiral models remains valid for spontaneous TSTs. As a result, intriguing merged TSTs profiles can form at small angles in the nonchiral Heisenberg and Heisenberg-IDMI models, like skyrmion - (trivial) MBs (Fig. 4b, d), antiskyrmion - (trivial) MBs (Fig. 4e), skyrmion – antiskyrmion pairs, and (anti)skyrmion clusters with high topological charges (Fig. 4a and Table 3). The $|Q| > 1$ spontaneous TSTs are allowed by the helicity degree of freedom in the nonchiral models. For example, the $Q = 3$ TST in Fig. 4a constitutes a bound state of three skyrmions with oppositely swirling spins. Such spontaneous skyrmionic clusters have been predicted only in itinerant magnets[39,81] and semiconducting $NiI_2$[40]. On the other hand, a sizeable NN DMI ($d = 0.15\ meV$) locks the chirality and induces spontaneous Néel-type skyrmions for $\theta \lesssim 3.15°$ in the chiral Heisenberg and Heisenberg-IDMI models (Table 3 and Supplementary Figure 8).

The spontaneous TSTs can be drastically tuned at $0\ K$. A magnetic field applied opposite to the core spins decouples the merged spin textures and confines them back to the AFM regions (Fig. 4b, c; Fig. 4e, f; Supplementary Movie 3). Conversely, the magnetization can be reversed by a magnetic field applied parallel to the core spins, trapping the reversed state's MBs in the AFM regions of the opposite layer (Supplementary Movie 4).

The merging and manipulation scenarios described above are also observed in the chiral/nonchiral Heisenberg-Kitaev models. The merging remains possible in these models throughout the inspected twist angle range $0.65° \leq \theta \leq 6°$. Moreover, the Kitaev interaction induces a canting-like effect for the spontaneous TSTs (Supplementary Figure 10), similar to their field-assisted counterparts.

Further insights on the spontaneous TSTs in all models are presented in Table 3. The results are selected from six simulations with distinct initial random states for each twist angle. We applied the same criterion as Table 2, with a preference for TSTs not trapped in the AFM regions.

Finally, we compare our methods and results with Akram *et al.*[59], who studied the TSTs in the chiral Heisenberg model, neglecting the thermal effects and adopting a continuum approximation of the bilayer Hamiltonian. The authors used a NN interlayer approach and performed LLG simulations at $0\ K$ starting from various initial FM states. The present work includes the thermal effects and uses an atomistic Hamiltonian. Further, we present an approach for the interlayer field beyond the NN and initiate the sLLG simulations from random spins. The differences in results and conclusions are summarized below.

Akram *et al.*[59] predicted three topological magnetic phases and concluded firm rules to determine the magnetic phase as a function of the twist angle. The first phase appears only at minimal angles, with Néel-type skyrmions confined to the AFM regions. However, in our study, simulations from various initial random states showed that this phase could be realized in the chiral Heisenberg model at any angle $0.65 < \theta \lesssim 3.15°$. For a short-range above the minimal angles, Akram *et al.*[59] reported a second phase with a single merged-skyrmion scenario. In particular, one skyrmion is formed over the three AFM regions in a layer, leaving the opposite layer FM. This merge avoids the FM regions and differs substantially from the various merged skyrmions reported in our study (Table 3). At relatively larger angles, the previous work[59] revealed a third magnetic phase with strips in one of the layers and a Néel-type skyrmion trapped in an AFM region of the opposite layer. In our work, the Néel-type skyrmion in this phase forms over a combination of AFM and FM regions. Moreover, we demonstrated that distinct initial states could generate merged skyrmions or magnetic strips at any angle $0.65° < \theta \lesssim 2.44°$, without firm rules. Further, we observed additional magnetic phases not captured previously, such as magnetic strips in both layers and skyrmions trapped in the FM regions.

**The twist-dependent averaged magnetization**

The sLLG approach offers valuable insights into the variation of the averaged magnetization with the temperature and the twist angle. Figures 5a and b present the $M - T$ curves for selected twist angles in the nonchiral Heisenberg and Heisenberg-IDMI models, respectively. The $M - T$ curves are comparable in the two models. The ordering temperature is virtually independent of the twist (Fig. 5a, b), with a value near $25\ K$. The ground state averaged magnetization (at $T = 0\ K$) varies smoothly with the twist, and the moiré bilayer approaches a pure FM ground state at large angles. This behavior follows the gradual alignment of the MB's spins along the positive z-axis in the nonchiral Heisenberg and Heisenberg-IDMI models. Above 4.3°, all spins acquire positive $S_z$ components (Fig. 5d, e) and the MBs disappear from the moiré superlattice. However, the normalized averaged magnetization remains slightly below unity (Fig. 5a, b) due to residual tilted spins near the cores of the AFM regions.

The Heisenberg-Kitaev model shows fundamentally different behavior. In particular, the MBs persist throughout the range $0.65° \leq \theta \leq 6°$, leaving the $M - T$ curve almost independent of the twist (Fig. 5c, f). Consequently, the ground state averaged magnetization does not approach the FM limit. Moreover, the Heisenberg-Kitaev model reveals a lower ordering temperature ($\sim 15K$). As a test, we added a sizable easy-axis anisotropy ($\mathcal{A} = 0.15\ meV$) to the Heisenberg-Kitaev model, and we did not observe significantly different results. We conclude that the Heisenberg-Kitaev model cannot reproduce the experimentally observed twist-dependent magnetic ground state[60].

**Discussion**

Technological implementation of TSTs crucially depends on discovering new topological magnetic materials and novel mechanisms for their stabilization. Moiré magnets are ideal candidates in this direction, which justifies the current tremendous interest in their fundamental and applied physics. Indeed, moiré skyrmionics is still in its early stages, promising vast opportunities for impactful discoveries. The recent experiments[60,61] on moiré $CrI_3$ constitute a significant advancement towards the experimental observation of TSTs in moiré magnets, which would require accurate mapping of the directions of spins in the moiré superlattice.

Theoretical models of moiré $CrI_3$ with sizeable NN DMI or dipolar interactions are expected to produce skyrmionic structures since these interactions are conventional sources for TSTs. Nevertheless, the dipolar interactions might be negligibly weak[55], while a significant NN DMI is not guaranteed and awaits future DFT or experimental confirmation. Consequently, it is essential to discover sources of topological magnetic textures that emerge exclusively from the moiré magnetism and go beyond the conventional skyrmion sources. This will pass through theoretical modeling that minimizes the approximations, conjugated with simulations that include the most relevant effects.

In our study, we investigated the ground state TSTs in moiré $CrI_3$, developing a study that accounts for the thermal effects, presents an atomistic approach for the interlayer coupling, and includes the IDM and Kitaev interactions on top of the Heisenberg in-plane exchange. We uncovered a stabilization mechanism for topological magnetic textures emerging from moiré interlayer fields with nontrivial textures. At large moiré periodicity, the whirling interlayer fields stabilize various types of ground state TSTs in the nonchiral models. Including the spin dynamics in both layers is crucial to account for this stabilization mechanism. Further, we showed that the monolayer approximation does not accurately describe the ground state manipulation.

The extra degrees of freedom (vorticity and helicity) characterizing the TSTs in nonchiral magnetic films attracted significant attention due to their technological relevance. These degrees of freedom are present in the nonchiral models of the moiré $CrI_3$, which allow the formation of skyrmionic clusters with high topological charges. A magnetic field is required temporarily during the cooling process to trap the TSTs in the AFM regions. Moreover, moiré $CrI_3$ broadens the class of materials that can host spontaneous TSTs. The moiré interlayer field constitutes a source for spontaneous TSTs, similar to the high-order spin interactions in itinerant magnets and the anisotropic exchange in $NiI_2$[40].

The Heisenberg model cannot describe the magnetic excitations in monolayer $CrI_3$. Accordingly, the pure Heisenberg model is not suitable to simulate the moiré magnetism in $CrI_3$. We have tested the Heisenberg-IDMI and the Heisenberg-Kitaev models and concluded that only the Heisenberg-IDMI is consistent with the twist-dependent ground state, observed experimentally[60] in moiré $CrI_3$. Therefore, our study suggests that the Heisenberg-IDMI is the correct model for $CrI_3$, in agreement with a very recent experimental study on topological magnons in monolayer $CrI_3$[82]. Note that our study does not account for the experimentally observed lattice relaxation at tiny twist angles $\theta \lesssim 0.5°$, which are excluded in the present study. Moreover, the lattice relaxation is found to be negligible[60] in experimental samples with $\theta \gtrsim 1°$. Further investigations are required to determine the relevance of lattice relaxation in the twist angle range $0.5° < \theta < 1°$.

Beyond $CrI_3$, our results suggest that Kitaev magnets might constitute better candidates for moiré skyrmionics than Heisenberg magnets. We demonstrated that a large Kitaev interaction could not dominate the stacking-dependent interlayer interaction. As a result, moiré Kitaev magnets can support nontrivial ground states over a broader twist angle range than Heisenberg magnets. These observations motivate attention towards Kitaev magnets for further advancement in moiré skyrmionics.

## References

1. Wang, X. S. & Wang, X. R. *Topology in Magnetism*. *Topics in Applied Physics* vol. 192 357–403 (Springer International Publishing, 2018).

2. Yu, G. *et al.* Room-temperature skyrmion shift device for memory application. *Nano Letters* **17**, 261–268 (2017).

3. Fert, A., Cros, V. & Sampaio, J. Skyrmions on the track. *Nature Nanotech* **8,** 152–156 (2013).

4. Sampaio, J., Cros, V., Rohart, S., Thiaville, A. & Fert, A. Nucleation, stability and current-induced motion of isolated magnetic skyrmions in nanostructures. *Nature Nanotech* **8**, 839–844 (2013).

5. Bogdanov, A. N. & Yablonskii, D. A. *Thermodynamically stable "vortices" in magnetically ordered crystals. The mixed state of magnets*, Sov. Phys. JETP 68, 101 (1989).

6. Mühlbauer, S. *et al.* Skyrmion lattice in a chiral magnet. *Science* **323**, 915–919 (2009).

7. Nayak, A. K. *et al.* Magnetic antiskyrmions above room temperature in tetragonal Heusler materials. *Nature* **548**, 561–566 (2017).

8. Yu, X. Z. *et al.* Biskyrmion states and their current-driven motion in a layered manganite. *Nat Commun* **5,** 3198 (2014).

9.  Göbel, B., Henk, J. & Mertig, I. Forming individual magnetic biskyrmions by merging two skyrmions in a centrosymmetric nanodisk. *Sci Rep* **9**, 9521 (2019).

10. Zhang, X. *et al.* Control and manipulation of a magnetic skyrmionium in nanostructures. *Phys. Rev. B* **94**, 094420 (2016).

11. Zhang, S., Kronast, F., van der Laan, G. & Hesjedal, T. Real-Space Observation of Skyrmionium in a Ferromagnet-Magnetic Topological Insulator Heterostructure. *Nano Letters* **18**, 1057–1063 (2018).

12. Gao, N. *et al.* Creation and annihilation of topological meron pairs in in-plane magnetized films. *Nat Commun* **10,** 5603 (2019).

13. Kharkov, Y. A., Sushkov, O. P. & Mostovoy, M. Bound States of Skyrmions and Merons near the Lifshitz Point. *Phys. Rev. Lett.* **119**, 207201 (2017).

14. Berg, B. & Lüscher, M. Definition and statistical distributions of a topological number in the lattice O(3) σ-model. *Nuclear Physics B* **190**, 412–424 (1981).

15. Yin, G. *et al.* Topological charge analysis of ultrafast single skyrmion creation. *Phys. Rev. B* **93**, 174403 (2016).

16. Belavin, A. A. & Polyakov, A. M. Metastable states of two-dimensional isotropic ferromagnets. *JETPL* **22**, 245 (1975).

17. Göbel, B., Mertig, I. & Tretiakov, O. A. Beyond skyrmions: Review and perspectives of alternative magnetic quasiparticles. *Physics Reports* **895**, 1–28 (2021).

18. Yu, X. Z. *et al.* Real-space observation of a two-dimensional skyrmion crystal. *Nature* **465**, 901–904 (2010).

19. Rößler, U. K., Bogdanov, A. N. & Pfleiderer, C. Spontaneous skyrmion ground states in magnetic metals. *Nature* ***442,** 797–801 (2006)*.

20. Yi, S. do, Onoda, S., Nagaosa, N. & Han, J. H. Skyrmions and anomalous Hall effect in a Dzyaloshinskii-Moriya spiral magnet. *Phys. Rev. B* **80**, 054416 (2009).

21. Heinze, S. *et al.* Spontaneous atomic-scale magnetic skyrmion lattice in two dimensions. *Nature Phys **7,** 713–718 (2011).*

22. Seki, S., Yu, X. Z., Ishiwata, S. & Tokura, Y. Observation of skyrmions in a multiferroic material. *Science* **336**, 198–201 (2012).

23. Peng, L. *et al.* Controlled transformation of skyrmions and antiskyrmions in a non-centrosymmetric magnet. *Nat. Nanotechnol. **15,** 181–186 (2020)*.

24. Yu, X. *et al.* Magnetic stripes and skyrmions with helicity reversals. *PNAS* **109**, 8856–8860 (2012).

25. Göbel, B., Henk, J. & Mertig, I. Forming individual magnetic biskyrmions by merging two skyrmions in a centrosymmetric nanodisk. *Sci Rep **9,** 9521 (2019)*.

26. Nagaosa, N. & Tokura, Y. Topological properties and dynamics of magnetic skyrmions. *Nature Nanotech* **8,** 899–911 (2013).

27. Okubo, T., Chung, S. & Kawamura, H. Multiple-q States and the Skyrmion Lattice of the Triangular-Lattice Heisenberg Antiferromagnet under Magnetic Fields. *Phys. Rev. Lett.* **108**, 017206 (2012).

28. Leonov, A. O. & Mostovoy, M. Multiply periodic states and isolated skyrmions in an anisotropic frustrated magnet. *Nature Communications* **6**, 8275 (2015).

29. Hayami, S., Lin, S.-Z. & Batista, C. D. Bubble and skyrmion crystals in frustrated magnets with easy-axis anisotropy. *Phys. Rev. B* **93**, 184413 (2016).

30. Zhang, X. *et al.* Skyrmion dynamics in a frustrated ferromagnetic film and current-induced helicity locking-unlocking transition. *Nature Communications* **8**, 1717 (2017).

31. Yu, X. Z. *et al.* Biskyrmion states and their current-driven motion in a layered manganite. *Nature Communications* **5**, 3198 (2014).

32. Wang, W. *et al.* A Centrosymmetric Hexagonal Magnet with Superstable Biskyrmion Magnetic Nanodomains in a Wide Temperature Range of 100-340 K. *Advanced Materials* **28**, 6887–6893 (2016).

33. Capic, D., Garanin, D. A. & Chudnovsky, E. M. Biskyrmion lattices in centrosymmetric magnetic films. *Physical Review Research* **1**, 033011 (2019).

34. Chudnovsky, E. M. & Garanin, D. A. Skyrmion glass in a 2D Heisenberg ferromagnet with quenched disorder. *New Journal of Physics* **20**, 033006 (2018).

35. Chudnovsky, E. M. & Garanin, D. A. Topological Order Generated by a Random Field in a 2D Exchange Model. *Phys. Rev. Lett.* **121**, 017201 (2018).

36. Roldán-Molina, A., Santander, M. J., Nunez, A. S. & Fernández-Rossier, J. Quantum fluctuations stabilize skyrmion textures. *Phys. Rev. B* **92**, 245436 (2015).

37. Martin, I. & Batista, C. D. Itinerant Electron-Driven Chiral Magnetic Ordering and Spontaneous Quantum Hall Effect in Triangular Lattice Models. *Phys. Rev. Lett.* **101**, 156402 (2008).

38. Akagi, Y. & Motome, Y. Spin Chirality Ordering and Anomalous Hall Effect in the Ferromagnetic Kondo Lattice Model on a Triangular Lattice. *Journal of the Physical Society of Japan* **79**, 083711 (2010).

39. Ozawa, R., Hayami, S. & Motome, Y. Zero-Field Skyrmions with a High Topological Number in Itinerant Magnets. *Phys. Rev. Lett.* **118**, 147205 (2017).

40. Amoroso, D., Barone, P. & Picozzi, S. Spontaneous skyrmionic lattice from anisotropic symmetric exchange in a Ni-halide monolayer. *Nature Communications* **11**, 5784 (2020).

41. Han, M. G. *et al.* Topological Magnetic-Spin Textures in Two-Dimensional van der Waals $Cr_2Ge_2Te_6$. *Nano Letters* **19**, 7859–7865 (2019).

42. Ding, B. *et al.* Observation of Magnetic Skyrmion Bubbles in a van der Waals Ferromagnet Fe3GeTe2. *Nano Letters* **20**, 868–873 (2020).

43. Yang, M. *et al.* Creation of skyrmions in van der Waals ferromagnet Fe3GeTe2 on (Co/Pd)n superlattice. *Sci. Adv.* **6,** eabb5157 (2020).

44. Wu, Y. *et al.* Néel-type skyrmion in WTe2/Fe3GeTe2 van der Waals heterostructure. *Nat Commun* **11,** 3860 (2020).

45. Xu, C. *et al.* Topological spin texture in Janus monolayers of the chromium trihalides Cr(I, X)3. *Phys. Rev. B* **101**, 060404 (2020).

46. Cui, Q., Liang, J., Shao, Z., Cui, P. & Yang, H. Strain-tunable ferromagnetism and chiral spin textures in two-dimensional Janus chromium dichalcogenides. *Phys. Rev. B* **102**, 094425 (2020).

47. Yuan, J. *et al.* Intrinsic skyrmions in monolayer Janus magnets. *Phys. Rev. B* **101**, 094420 (2020).

48. Sun, W. *et al.* Controlling bimerons as skyrmion analogues by ferroelectric polarization in 2D van der Waals multiferroic heterostructures. *Nat Commun* **11,** 5930 (2020).

49. Xu, C. *et al.* Electric-Field Switching of Magnetic Topological Charge in Type-I Multiferroics. *Phys. Rev. Lett.* **125**, 037203 (2020).

50. Liang, J., Cui, Q. & Yang, H. Electrically switchable Rashba-type Dzyaloshinskii-Moriya interaction and skyrmion in two-dimensional magnetoelectric multiferroics. *Phys. Rev. B* **102**, 220409 (2020).

51. Lu, X., Fei, R., Zhu, L. & Yang, L. Meron-like topological spin defects in monolayer CrCl3. *Nat Commun* **11,** 4724 (2020).

52. Augustin, M., Jenkins, S., Evans, R. F. L., Novoselov, K. S. & Santos, E. J. G. Properties and dynamics of meron topological spin textures in the two-dimensional magnet CrCl3. *Nature Communications* **12**, 185 (2021).

53. Tong, Q., Liu, F., Xiao, J. & Yao, W. Skyrmions in the Moiré of van der Waals 2D Magnets. *Nano Letters* **18**, 7194–7199 (2018).

54. Hejazi, K., Luo, Z. X. & Balents, L. Noncollinear phases in moiré magnets. *PNAS* **117**, 10721–10726 (2020).

55. Xiao, F., Chen, K. & Tong, Q. Magnetization textures in twisted bilayer Cr X 3 ( X =Br, I) . *Physical Review Research* **3**, 013027 (2021).

56. Akram, M. & Erten, O. Skyrmions in twisted van der Waals magnets. *Phys. Rev. B* **103**, L140406 (2021).

57. Hejazi, K., Luo, Z.-X. & Balents, L. Heterobilayer moiré magnets: moiré skyrmions and the commensurate-incommensurate transition. Phys. Rev. B **104**, L100406 (2021).

58. Ray, S. & Das, T. Hierarchy of multi-order skyrmion phases in twisted magnetic bilayers. *Phys. Rev. B* **104**, 014410 (2021).

59. Akram, M. *et al.* Moiré Skyrmions and Chiral Magnetic Phases in Twisted CrX3(X = I, Br, and Cl) Bilayers. *Nano Letters* **21**, 6633–6639 (2021).

60. Xu, Y. *et al.* Coexisting ferromagnetic–antiferromagnetic state in twisted bilayer CrI3. *Nat. Nanotechnol.* **17, 143–147 (2022)**.

61. Song, T. *et al.* Direct visualization of magnetic domains and moiré magnetism in twisted 2D magnets. *Science* **374**, 1140–1144 (2021).

62. Lee, I. *et al.* Fundamental Spin Interactions Underlying the Magnetic Anisotropy in the Kitaev Ferromagnet CrI3. *Phys. Rev. Lett.* **124**, 017201 (2020).

63. Chen, L. *et al.* Magnetic anisotropy in ferromagnetic CrI3. *Phys. Rev. B* **101**, 134418 (2020).

64. Chen, L. *et al.* Topological Spin Excitations in Honeycomb Ferromagnet CrI3. *Physical Review X* **8**, 041028 (2018).

65. Soriano, D., Katsnelson, M. I. & Fernández-Rossier, J. Magnetic Two-Dimensional Chromium Trihalides: A Theoretical Perspective. *Nano Letters* **20**, 6225–6234 (2020).

66. Dzyaloshinsky, I. A thermodynamic theory of "weak" ferromagnetism of antiferromagnetics. *Journal of Physics and Chemistry of Solids* **4**, 241–255 (1958).

67. Moriya, T. Anisotropic superexchange interaction and weak ferromagnetism. *Physical Review* **120**, 91–98 (1960).

68. Owerre, S. A. A first theoretical realization of honeycomb topological magnon insulator. *Journal of Physics Condensed Matter* **28**, 386001 (2016).

69. Ghader, D. Insights on magnon topology and valley-polarization in 2D bilayer quantum magnets. *New Journal of Physics* **23**, 053022 (2021).

70. Kitaev, A. Anyons in an exactly solved model and beyond. *Annals of Physics* **321**, 2–111 (2006).

71. Xu, C., Feng, J., Xiang, H. & Bellaiche, L. Interplay between Kitaev interaction and single ion anisotropy in ferromagnetic CrI3 and CrGeTe3 monolayers. *npj Computational Materials* **4**, 57 (2018).

72. Li, T. *et al.* Pressure-controlled interlayer magnetism in atomically thin CrI3. *Nature Materials* **18**, 1303–1308 (2019).

73. Song, T. *et al.* Switching 2D magnetic states via pressure tuning of layer stacking. *Nature Materials* **18**, 1298–1302 (2019).

74. Sivadas, N., Okamoto, S., Xu, X., Fennie, C. J. & Xiao, D. Stacking-Dependent Magnetism in Bilayer CrI 3. *Nano Letters* **18**, 7658–7664 (2018).

75. Jiang, P. *et al.* Stacking tunable interlayer magnetism in bilayer CrI3. *Phys. Rev. B* **99**, 144401 (2019).

76. Li, S. *et al.* Magnetic-Field-Induced Quantum Phase Transitions in a van der Waals Magnet. *Physical Review X* **10**, 011075 (2020).

77. Evans, R. F. L. *et al.* Atomistic spin model simulations of magnetic nanomaterials. *Journal of Physics Condensed Matter,* **26**, 103202 (2014).

78. Wolfram Mathematica: Modern Technical Computing. https://www.wolfram.com/mathematica/.

79. García-Palacios, J. L. & Lázaro, F. J. Langevin-dynamics study of the dynamical properties of small magnetic particles. *Phys. Rev. B - Condensed Matter and Materials Physics* **58**, 14937–14958 (1998).

80. Jena, J. *et al.* Elliptical Bloch skyrmion chiral twins in an antiskyrmion system. *Nat Commun* **11, 1115 (2020)**.

81. Hayami, S. & Motome, Y. Effect of magnetic anisotropy on skyrmions with a high topological number in itinerant magnets. *Phys. Rev. B* **99**, 094420 (2019).

82. Chen, L. *et al.* Magnetic field effect on topological spin excitations in CrI3. *Physical Review X* **11**, 031047 (2021).

**Acknowledgments**

Part of the numerical calculations was performed using the Phoenix High Performance Computing facility at the American University of the Middle East (AUM), Kuwait. D. G. thanks Qingjun Tong for sharing the DFT results in Figure 1b.

**Author contributions**

D.G. planned the research with inputs from A.S. and B.J. D.G. did the analytic calculations, performed the Mathematica numerical calculations, and drafted the manuscript. D.G. analyzed the results with inputs from A.S. and B.J. B.J. performed the Vampire numerical simulations. B.J. and D.G. prepared the Supplementary Information. All authors revised the manuscript.

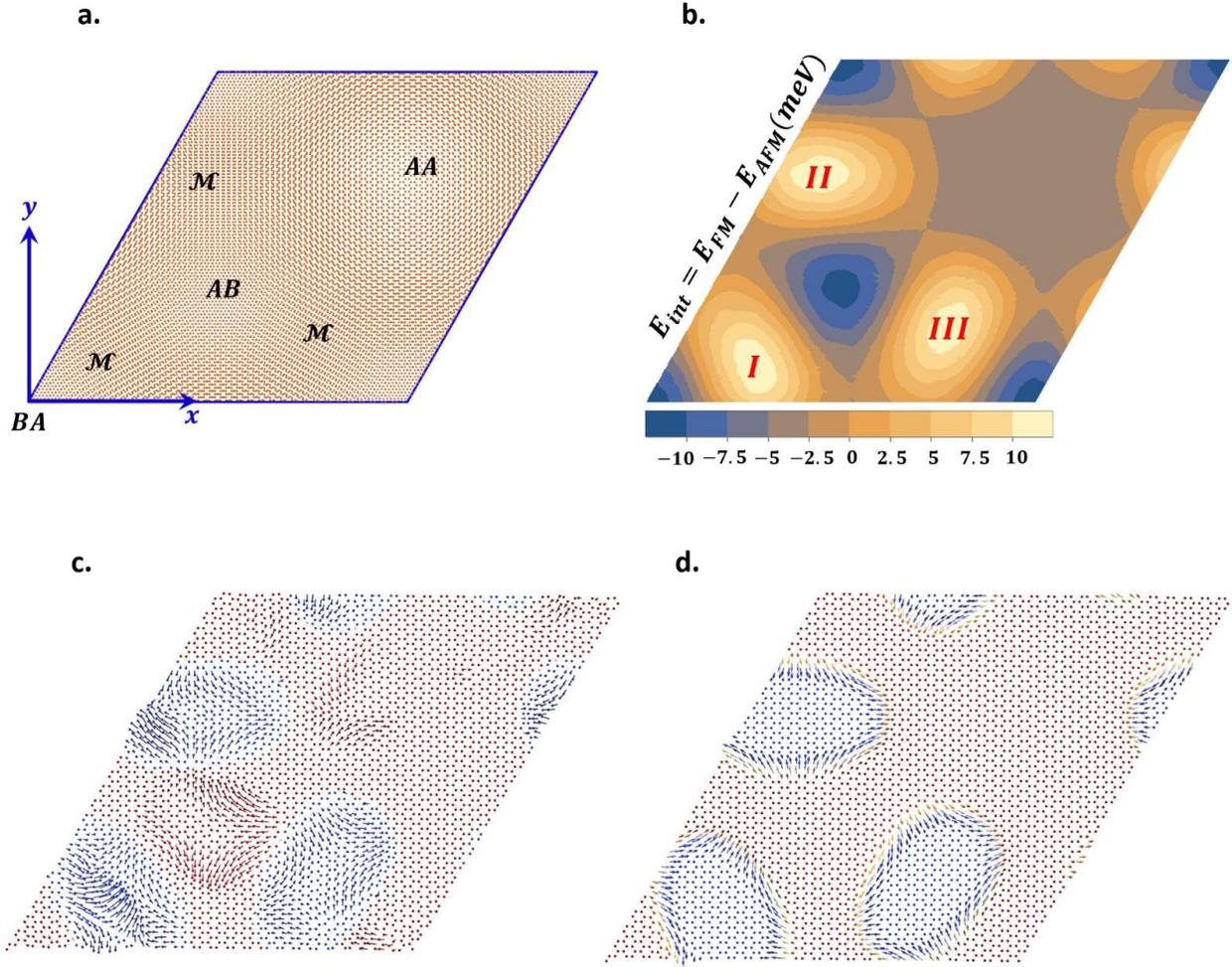

**Fig. 1 Antivortex moiré interlayer field stabilizing ground state antiskyrmions in the nonchiral Heisenberg model. a** Rhombohedral (BA, AB, and AA) and monoclinic ($\mathcal{M}$) stacking regions in the $CrI_3$ moiré supercell for a twist angle $\theta = 0.96°$. **b** The interlayer exchange energy ($E_{int}$) as a function of the interlayer translation in a moiré supercell. The plot is reproduced from the density functional theory data shared by the authors of a previous work[55]. The interlayer coupling is antiferromagnetic in the monoclinic regions (I, II, and III). **c, d** The normalized moiré interlayer field (**c**) and antiskyrmions at $0K$ (d), obtained in the Heisenberg model for twist angle $\theta = 1.35°$ and external magnetic field strength $B = 1T$. The interlayer field displays antivortex textures at the antiferromagnetic regions and stabilizes the antiskyrmions. The vectors are presented using a temperature-color map which indicates the out-of-plane components ranging from $-1$ (blue) to $1$ (red).

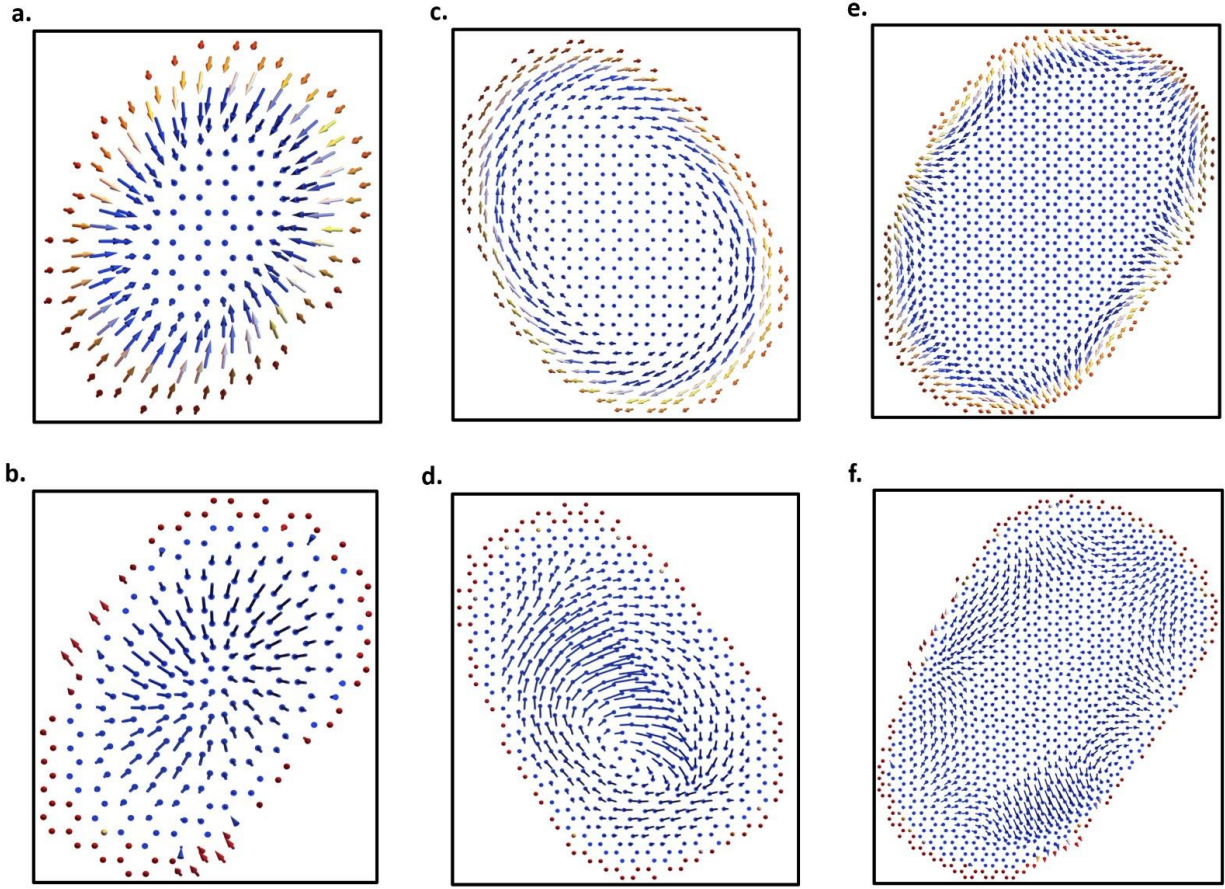

**Fig. 2 Multi-flavor ground state topological spin textures in the nonchiral Heisenberg and Heisenberg-IDMI models. a, b** Néel-type skyrmion (**a**) stabilized by an interlayer field with a radial vortex profile (**b**). The skyrmion forms in region III of the Heisenberg model with twist angle $\theta = 2°$ and external magnetic field strength $B = 1T$. **c, d** Bloch-type skyrmion (**c**) stabilized by an interlayer field with a sink vortex profile (**d**). The skyrmion forms in region I of the Heisenberg-IDMI model with $\theta = 1.2°$ and $B = 1.25T$. **e, f** Antiskyrmion pair with topological charge $Q = 2$ (**e**) and the corresponding interlayer field (**f**) in region I of the Heisenberg-IDMI model with $\theta = 0.76°$ and $B = 1.5T$. Note that IDMI stands for the intrinsic Dzyaloshinskii-Moriya interaction.

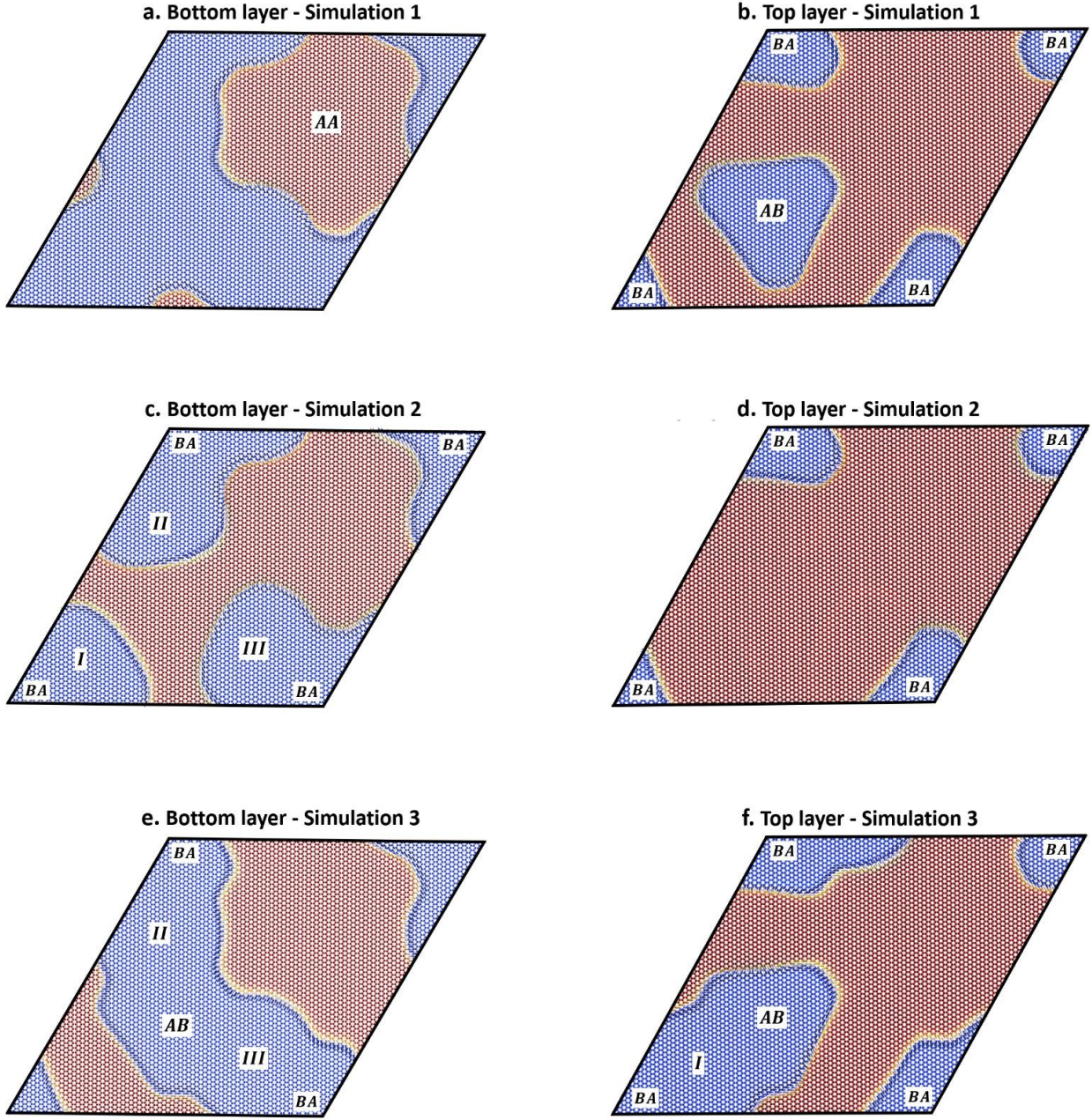

**Fig. 3 Dependence of the spontaneous topological spin textures (TSTs) on the initial spin configuration.** An illustrative example of distinct topological ground states obtained from different initial random configurations in the Heisenberg model with twist angle $\theta = 0.89°$ and external magnetic field strength $B = 0\ T$. Simulation 1 results in a TST with topological charge $Q = -2$ in the ferromagnetic (FM) AA region of the bottom layer (**a**). Meanwhile, the top layer (**b**) hosts a trivial magnetic bubble in the FM BA region and a TST ($Q = -1$) in the FM AB region. In simulation 2, the bottom layer (**c**) displays a single TST ($Q = 1$) that extends over the monoclinic regions I, II, III, and the FM region BA. The top layer (**d**) hosts a TST ($Q = 1$) in the BA region. Simulation 3 results in a magnetic strip in the bottom layer (**e**) and a TST ($Q = -2$) combining FM regions AB, BA, and the monoclinic region I (**f**). To better visualize the reported TSTs, we present the same results duplicated over a $2 \times 2$ moiré superlattice in Supplementary Figure 9.

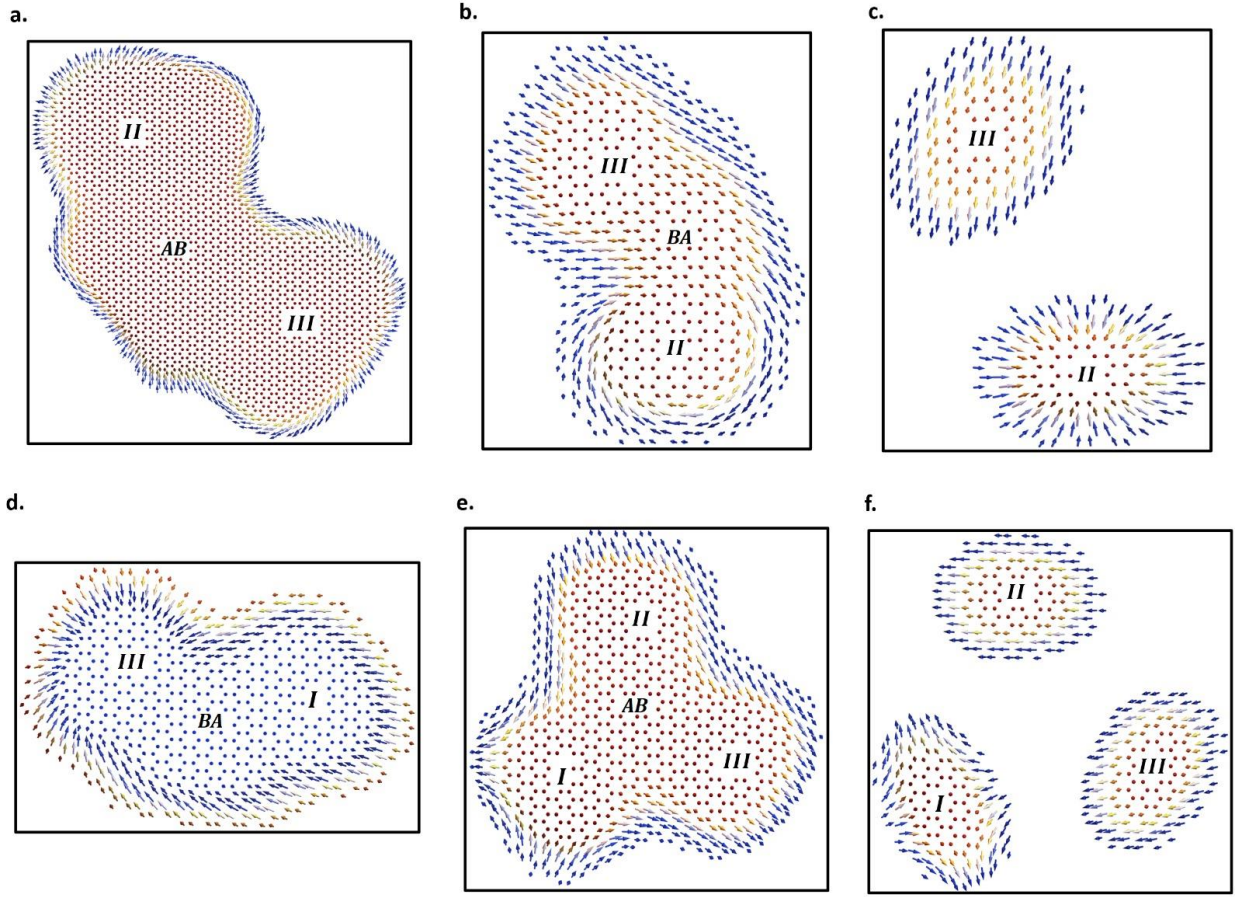

**Fig. 4 Merged spontaneous topological spin textures (TSTs) in the nonchiral Heisenberg and Heisenberg-IDMI models. a** TSTs in the bottom layer of the nonchiral Heisenberg-IDMI model for twist angle $\theta = 0.96°$ (IDMI stands for the intrinsic Dzyaloshinskii-Moriya interaction). The TST extends over regions II, III, and the AB-stacked ferromagnetic (FM) region. It hosts more than 2800 spins, carries a topological charge $Q = 3$, and represents a bound state of three skyrmions with different helicities. **b** The nonchiral Heisenberg-IDMI model with $\theta = 2.28°$ hosts a TST combining a Bloch-type skyrmion (region II) and a topologically trivial magnetic bubble (MB) in region III. Applying a magnetic field $-0.5\hat{z}T$ at $0\ K$ separates the MB and the skyrmion (**c**) while changing the skyrmion's morphology to a Néel-type skyrmion. **d, e** TSTs in the Heisenberg model for $\theta = 2°$ (**d**) and $\theta = 2.13°$ (**e**). They correspond to a Néel-type skyrmion merged with a topologically trivial MB (**d**), and an antiskyrmion merged with two topologically trivial MBs (**e**). **f** The merged spin configuration in (**e**) can be detached by a magnetic field $-0.5\hat{z}T$ at $0\ K$. The decoupling inverts the helicity of the antiskyrmion confined to region I.

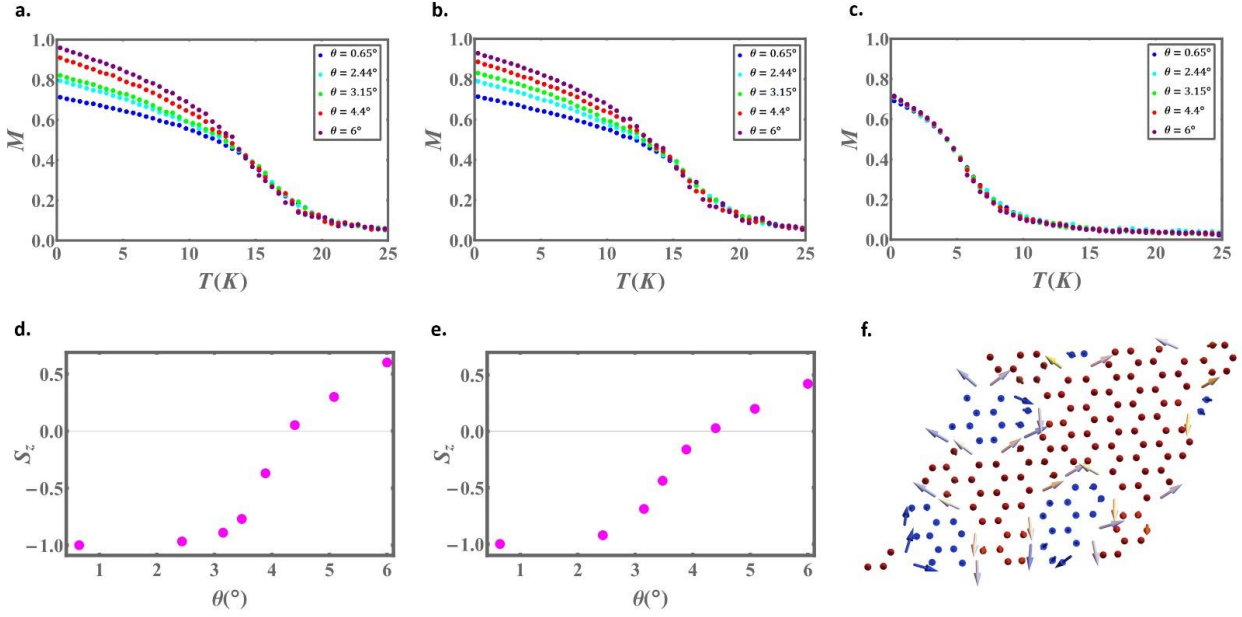

**Fig. 5 Twist angle effect on the ground state averaged magnetization ($M$). a, b, c** The $M - T$ curves ($T$ represents temperature) for selected twist angles in the nonchiral Heisenberg (**a**), Heisenberg-IDMI (**b**), and Heisenberg-Kitaev (**c**) models. IDMI stands for the intrinsic Dzyaloshinskii-Moriya interaction. The twisted bilayers are polarized by a magnetic field $\boldsymbol{B} = 1\hat{\boldsymbol{z}}T$. In (**a**) and (**b**), the averaged magnetization approaches the pure FM ground state in the large twist angle interval $\theta \gtrsim 4.3°$. However, in (**c**), the averaged magnetization is almost independent of the twist angle. **d, e** The minimal $S_z$ ($z$ component of the spin) value in the moiré supercell as a function of the twist angle for the Heisenberg (**d**) and the Heisenberg-IDMI (**e**) models at $T = 0K$. A steep increase in $S_z$ is observed in the interval $\theta \gtrsim 3°$. At $\theta \gtrsim 4.3°$, the magnetic bubbles (MBs) disappear in both models. **f** Persistent ground state MBs in the top layer of the nonchiral Heisenberg-Kitaev model at $\theta = 6°$.

| Model | Parameters |
|---|---|
| Heisenberg Model[60] | $J = 2\ meV$, $\mathcal{A} = 0.3\ meV$, $K = \Gamma = D = d = 0\ meV$. |
| Heisenberg-IDMI Model[63] | $J = 2.13\ meV$, $\mathcal{A} = 0.2\ meV$, $D = 0.193\ meV$, $K = \Gamma = d = 0\ meV$. |
| Heisenberg-Kitaev Model[62] | $J = 0.2\ meV$, $K = 5.2\ meV$, $\Gamma = 0.0675\ meV$, $\mathcal{A} = D = d = 0\ meV$. |

**Table 1. Models' parameters.** Numerical values for the Hamiltonian's parameters in the Heisenberg, Heisenberg-IDMI, and Heisenberg-Kitaev models for a monolayer $CrI_3$. IDMI stands for the intrinsic Dzyaloshinskii-Moriya interaction with strength $D$. The parameters $J$ and $\mathcal{A}$ are the intralayer Heisenberg coupling and the easy axis magnetic anisotropy, respectively. $K$ and $\Gamma$ are parameters that describe the Kitaev interaction. We note that the nearest-neighbor Dzyaloshinskii-Moriya interaction parameter $d$ is zero in the monolayer case due to the centrosymmetric symmetry of the honeycomb lattice.

| Model | Magnetic Field ($T$) | $\theta = 0.65°$ | $\theta = 0.86°$ | $\theta = 1.02°$ | $\theta = 1.35°$ | $\theta = 2.13°$ |
|---|---|---|---|---|---|---|
| Heisenberg | $B = 0.25$ | $Q_I^B = 1*$<br>$Q_{II}^B = 1*$<br>$Q_{III}^T = -1**$ | $Q_I^B = -1**$<br>$Q_{II}^B = -1**$<br>$Q_{III}^B = 0$ | $Q_I^T = 0$<br>$Q_{II}^T = -1**$<br>$Q_{III}^B = 1*$ | $Q_I^B = -1**$<br>$Q_{II}^B = 1*$<br>$Q_{III}^B = -1\bullet$ | – |
| | $B = 0.5$ | $Q_I^B = 1*$<br>$Q_{II}^T = 1*$<br>$Q_{III}^T = 0$ | $Q_I^T = 0$<br>$Q_{II}^T = -2\bullet*$<br>$Q_{III}^T = 0$ | $Q_I^T = -1**$<br>$Q_{II}^T = -1**$<br>$Q_{III}^B = 1*$ | $Q_I^T = 0$<br>$Q_{II}^T = -1**$<br>$Q_{III}^B = -1\bullet$ | $Q_I^T = 0$<br>$Q_{II}^T = 0$<br>$Q_{III}^T = 1*$ |
| | $B = 1$ | $Q_I^B = -1\bullet$<br>$Q_{II}^B = 0$<br>$Q_{III}^B = -1**$ | $Q_I^B = -1**$<br>$Q_{II}^B = -1\bullet$<br>$Q_{III}^B = 1*$ | $Q_I^T = 1*$<br>$Q_{II}^T = -1**$<br>$Q_{III}^B = 1*$ | $Q_I^B = 1*$<br>$Q_{II}^B = 1*$<br>$Q_{III}^B = 1*$ | $Q_I^T = 0$<br>$Q_{II}^T = 1*$<br>$Q_{III}^T = 1*$ |
| Heisenberg-IDMI | $B = 0.25$ | $Q_I^B = 1*$<br>$Q_{II}^B = -1\bullet$<br>$Q_{III}^B = -1**$ | $Q_I^T = 1*$<br>$Q_{II}^B = 0$<br>$Q_{III}^B = 1*$ | $Q_I^T = 0$<br>$Q_{II}^T = -1**$<br>$Q_{III}^T = 1*$ | $Q_I^T = 0$<br>$Q_{II}^T = -1**$<br>$Q_{III}^B = 1*$ | – |
| | $B = 0.5$ | $Q_I^B = 1*$<br>$Q_{II}^B = -1\bullet$<br>$Q_{III}^T = -1\bullet$ | $Q_I^T = 1*$<br>$Q_{II}^B = -1\bullet$<br>$Q_{III}^B = 0$ | $Q_I^T = 0$<br>$Q_{II}^T = -1**$<br>$Q_{III}^B = 0$ | $Q_I^B = 0$<br>$Q_{II}^T = 0$<br>$Q_{III}^B = 1*$ | $Q_I^T = 0$<br>$Q_{II}^T = 1*$<br>$Q_{III}^T = 0$ |
| | $B = 1$ | $Q_I^B = 2\bullet\bullet$<br>$Q_{II}^B = -1\bullet$<br>$Q_{III}^T = 1*$ | $Q_I^T = 1*$<br>$Q_{II}^T = 1*$<br>$Q_{III}^B = 1*$ | $Q_I^T = 0$<br>$Q_{II}^T = 0$<br>$Q_{III}^B = -1**$ | $Q_I^B = 1*$<br>$Q_{II}^T = 1*$<br>$Q_{III}^B = 1*$ | – |
| Heisenberg-Kitaev | $B = 0.25$ | $Q_I^B = 0$<br>$Q_{II}^T = 1$<br>$Q_{III}^B = -1$ | $Q_I^B = -1$<br>$Q_{II}^T = 1$<br>$Q_{III}^T = 1$ | $Q_I^T = 0$<br>$Q_{II}^B = 2$<br>$Q_{III}^B = 0$ | $Q_I^B = -1$<br>$Q_{II}^T = 1$<br>$Q_{III}^B = -1$ | $Q_I^B = 1$<br>$Q_{II}^T = -1$<br>$Q_{III}^B = 1$ |
| | $B = 0.5$ | $Q_I^B = -1$<br>$Q_{II}^B = -1$<br>$Q_{III}^T = 1$ | $Q_I^B = -2$<br>$Q_{II}^T = 1$<br>$Q_{III}^T = 1$ | $Q_I^B = -1$<br>$Q_{II}^B = 1$<br>$Q_{III}^B = -1$ | $Q_I^T = 1$<br>$Q_{II}^B = -1$<br>$Q_{III}^B = 1$ | $Q_I^T = 1$<br>$Q_{II}^B = -1$<br>$Q_{III}^B = 1$ |
| | $B = 1$ | $Q_I^B = -1$<br>$Q_{II}^T = -1$<br>$Q_{III}^T = 1$ | $Q_I^T = 1$<br>$Q_{II}^T = 1$<br>$Q_{III}^T = -1$ | $Q_I^B = 0$<br>$Q_{II}^B = -1$<br>$Q_{III}^B = -1$ | $Q_I^T = 1$<br>$Q_{II}^T = 1$<br>$Q_{III}^T = 1$ | $Q_I^T = 1$<br>$Q_{II}^B = -1$<br>$Q_{III}^T = 1$ |

**Table 2. Extended results on the field-assisted topological spin textures (TSTs).** The table elaborates the rich spectrum of field-assisted TSTs in the nonchiral models, presenting selected results from several simulations with distinct random initial states as described in the text. The symbols $\theta$ and $B$ stand for the twist angle and the magnitude of the external magnetic field, respectively. We have calculated the topological charge $Q$ from Equation (1), using the numerical algorithm detailed by M. Augustin *et al.*[52] (see their supplementary materials for full details). The subscript of $Q$ specifies the antiferromagnetic region to which the TST is confined. The superscripts $B$ and $T$ stand for the bottom ($l = 1$) and top ($l = 2$) layers, respectively. IDMI stands for the intrinsic Dzyaloshinskii-Moriya interaction. In the Heisenberg and the Heisenberg-IDMI models, the symbol next to the topological charge indicates the TST's nature. We used the following conventions: antiskyrmion*, Bloch-type skyrmion**, Néel-type skyrmion•, antiskyrmion pair••, and skyrmion pair•*. In the Heisenberg-Kitaev model, the TSTs are identified as vortices ($Q < 0$) and antivortices ($Q > 0$). The dashes indicate the absence of TSTs.

| Model | $\theta = 0.65°$ | $\theta = 0.86°$ | $\theta = 1.02°$ | $\theta = 1.35°$ | $\theta = 2.13°$ |
|---|---|---|---|---|---|
| **Nonchiral Heisenberg** | $Q^B_{III,AB} = -1$ $Q^T_{I,II,BA} = 3$ | $Q^B_{I-III,BA} = 1$ $Q^T_{BA} = 1$ | $Q^B_{I-III,AB} = 1$ $Q^T_{AB} = 1$ | $Q^B_{BA} = 1$ $Q^T_{I-III,BA} = -1$ | $Q^B_{I-III,AB} = -1$ $Q^T_{AB} = 0$ |
| **Chiral Heisenberg** | $Q^B_{II} = 1$ $Q^B_{III} = 1$ $Q^T_I = 1$ | $Q^B_{I-III,AB} = -1$ $Q^T_{AB} = 0$ | $Q^B_{II,III,AB} = 1$ $Q^T_{I,AB} = 1$ | $Q^B_{I,BA} = -1$ $Q^T_{BA,II,III} = -1$ | $Q^B_{BA} = -1$ $Q^T_{I-III,BA} = -1$ |
| **Nonchiral Heisenberg-IDMI** | $Q^B_{I-III,AB} = 1$ $Q^T_{AB} = -1$ | Magnetic strips in $l = 1$ and $Q^T_{BA,I,AB} = -2$ | $Q^B_I = 1$ $Q^B_{II} = 0$ $Q^B_{III} = 1$ | $Q^B_{I,BA} = 0$ $Q^T_{II,BA,III} = 1$ | Magnetic strips in $l = 1$ and $Q^T_{BA,III,AB} = 1$ |
| **Chiral Heisenberg-IDMI** | $Q^B_{I-III,AB} = 1$ $Q^T_{AB} = 1$ | $Q^B_{III,AB} = -1$ $Q^T_{I,II,AB} = -1$ | $Q^B_{II,AB,III} = 1$ $Q^T_{I,AB} = 1$ | $Q^B_{I,BA} = -1$ $Q^T_{II,BA,III} = -1$ | $Q^B_{BA} = -1$ $Q^T_{I-III,BA} = -1$ |
| **Nonchiral Heisenberg-Kitaev** | $Q^B_{BA,AA} = -1$ $Q^T_{AB,AA} = -2$ | $Q^B_{II,BA,III} = 1$ $Q^T_{I,BA} = 0$ | $Q^B_{I,AB,III} = 1$ $Q^T_{I,AB} = -1$ | $Q^B_{I,III,BA} = -1$ $Q^T_{BA,II} = 1$ | $Q^B_{III,AA} = 0$ $Q^B_{AB,III,BA} = -1$ |
| **Chiral Heisenberg-Kitaev** | $Q^B_{AB} = -1$ $Q^B_{BA} = -1$ $Q^T_{AA} = -1$ | $Q^B_{AB,III,BA} = 1$ and Magnetic strips in $l = 2$ | $Q^B_{I,AB} = 1$ $Q^T_{II,AB,III} = 1$ | $Q^B_{I,II,BA} = 1$ $Q^T_{III,BA} = 1$ | $Q^B_{II,III,AB} = 1$ $Q^T_{I,AB} = 1$ |

**Table 3. Extended results on the spontaneous topological spin textures (TSTs).** Selected spontaneous ground state TSTs in the chiral and nonchiral versions of the three models. The criterion for selecting the results is described in the text. IDMI stands for the intrinsic Dzyaloshinskii-Moriya interaction. The subscript of the topological charge $Q$ specifies the (merged) regions over which the TST is localized. $I - III$ stands for the monoclinic regions $I, II$, and $III$ of the moiré supercell.

# Supplementary Figures
# Whirling interlayer fields as a source of stable topological order in moiré CrI$_3$

Doried Ghader, Bilal Jabakhanji, and Alessandro Stroppa

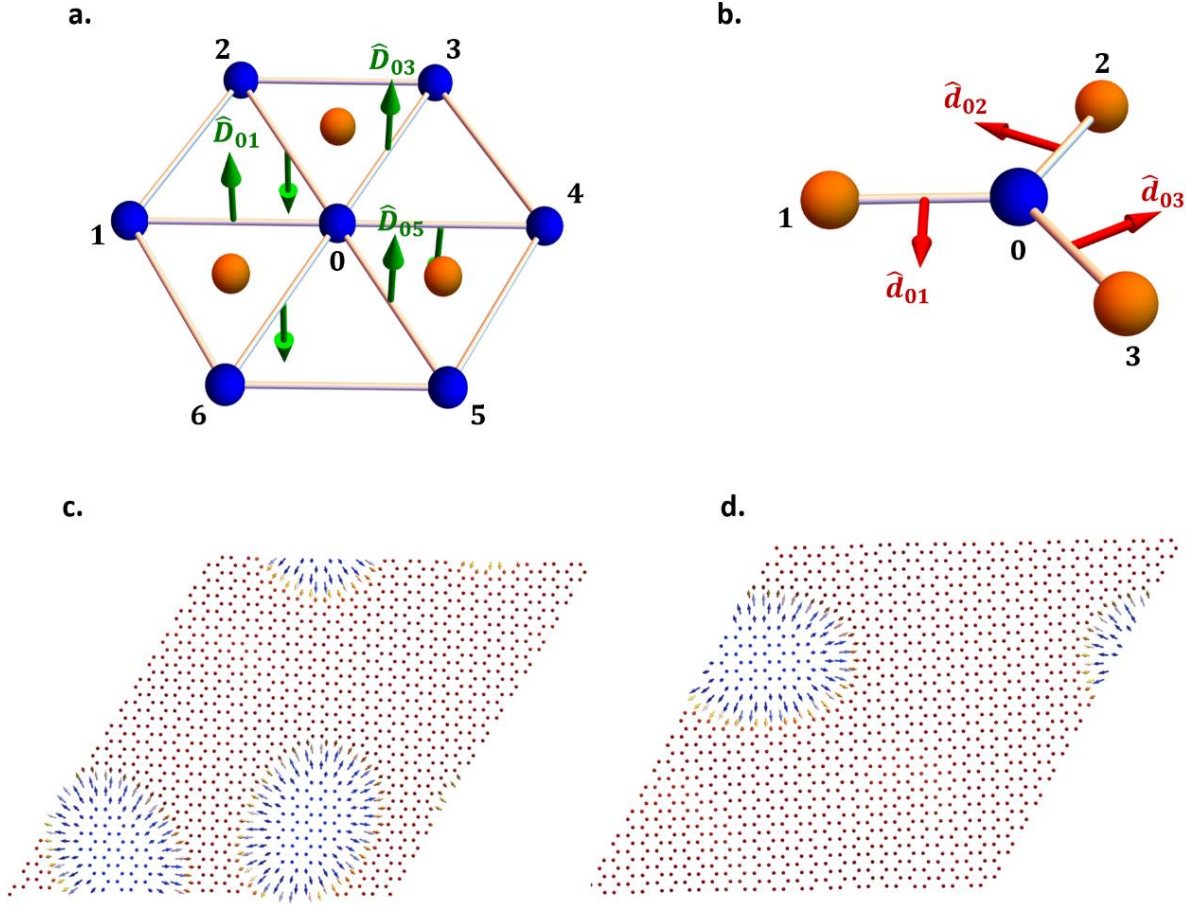

**Supplementary Figure 1. a** The next NN DMI (green arrows) oriented along $\pm\hat{z}$. This Néel-type DMI can explain the topological gaps observed in the magnon spectrum of monolayer $CrI_3$. **b** The in-plane chiral DMI (red arrows) between NN atoms. The chiral DMI is absent in monolayer $CrI_3$ but can emerge in the twisted phase due to the broken centrosymmetry in the moiré superlattice. Sizeable NN DMI locks the TSTs' vorticity in the Heisenberg and the Heisenberg-IDMI models to produce Néel-type skyrmions. **c, d** Sample of the Néel-type skyrmions stabilized by the NN DMI (0.15 $meV$) in the bottom (**c**) and top (**d**) layers of the Heisenberg-IDMI model with $\theta = 2.45°$ and $B = 0.5\ T$. The vectors are presented using a temperature-color map which indicates the out-of-plane components ranging from -1 (blue) to 1 (red).

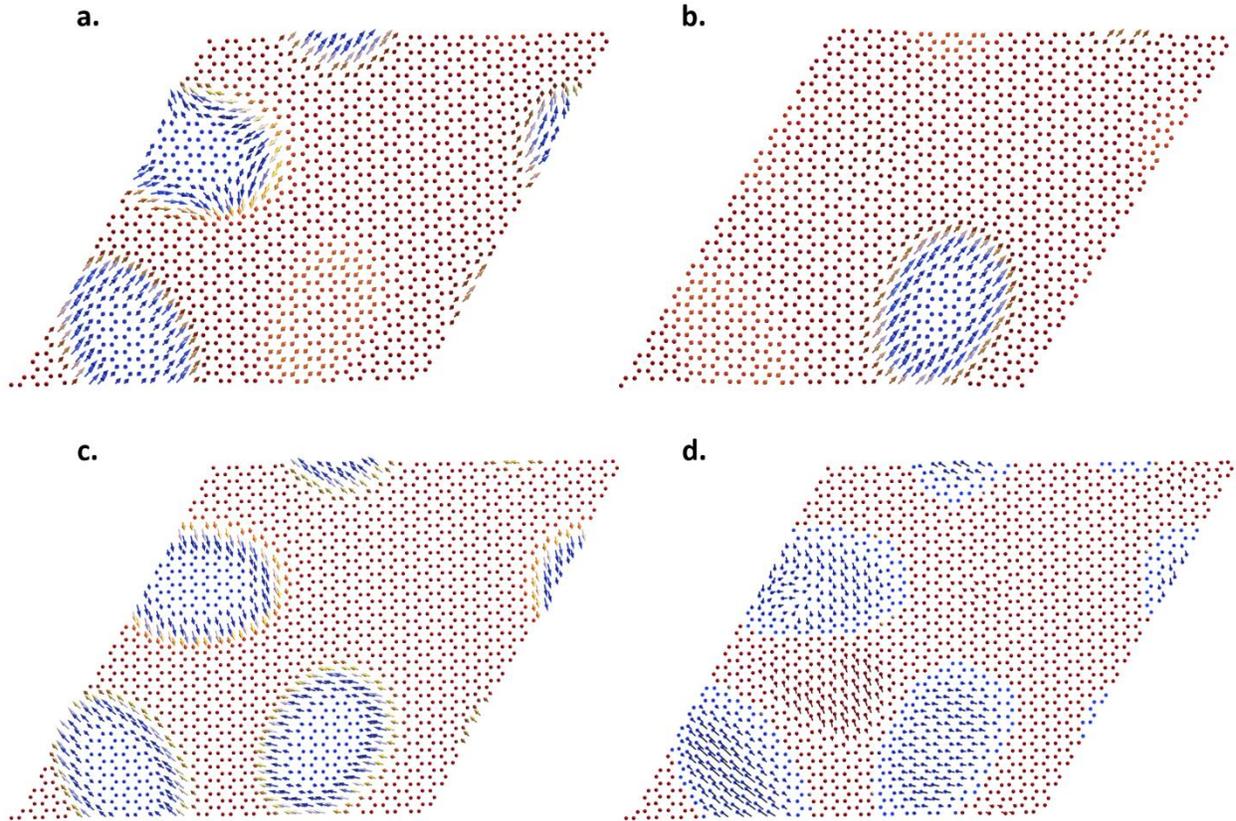

**Supplementary Figure 2. a, b** Field-assisted spin textures in the top (**a**) and bottom (**b**) layers of the nonchiral Heisenberg-IDMI model with $\theta = 2.13°$ and $B = 0.5\ T$. The top layer hosts an antiskyrmion, while the rest of the MBs are trivial. **c, d** Nontopological MBs (**c**) stabilized by interlayer moiré fields with trivial textures (**d**). The results correspond to the top layer of the nonchiral Heisenberg model with $\theta = 1.89°$ and $B = 1.25\ T$.

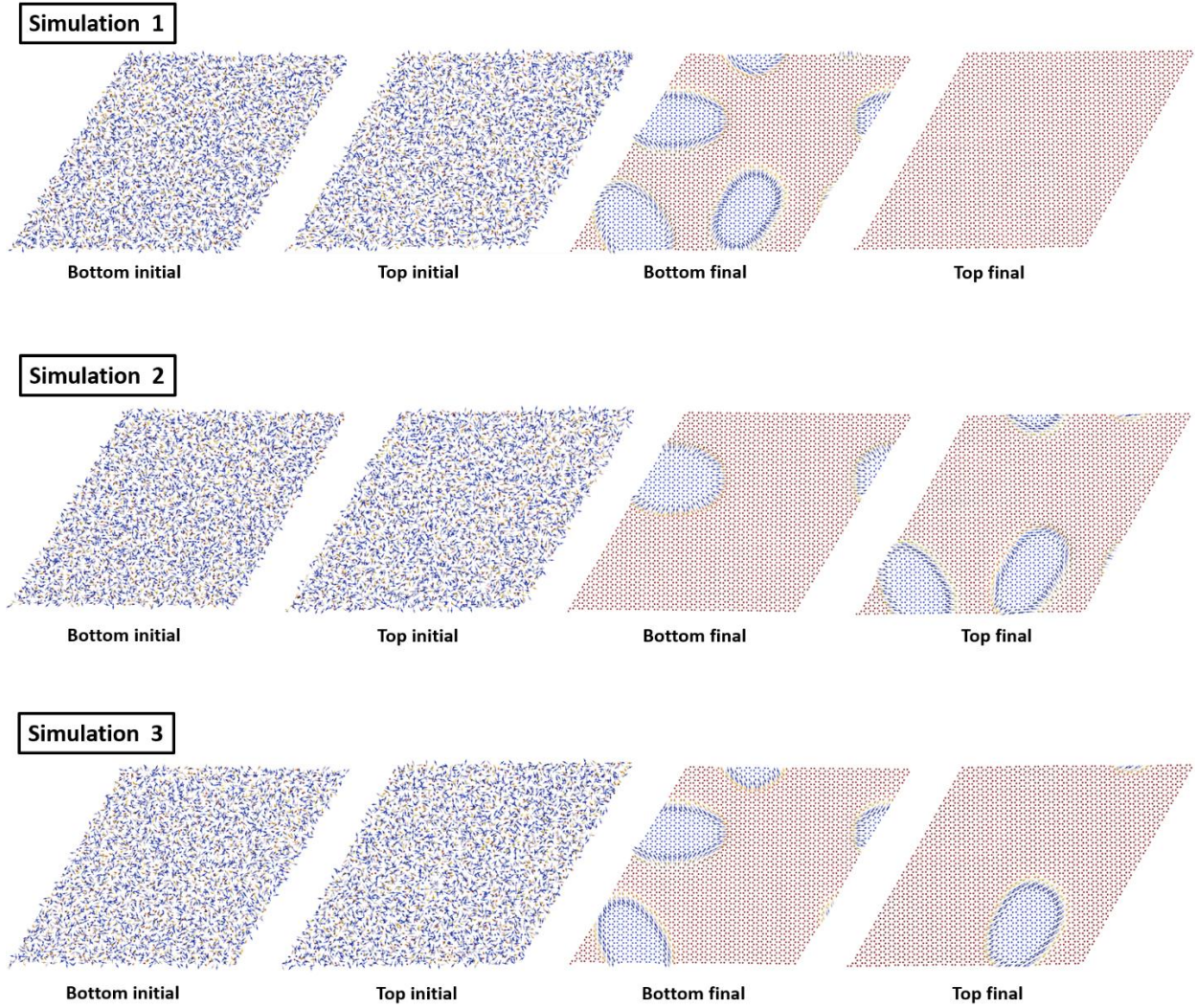

**Supplementary Figure 3.** Demonstration of the crucial dependence of the ground state on the random initial spin configuration. The results correspond to the Heisenberg model with $\theta = 1.35°$ and $B = 0.25\ T$. Distinct random spin configurations are used in simulations 1, 2, and 3, resulting in different ground states, with ($Q_I^B = -1^{**}$, $Q_{II}^B = 1^*$, $Q_{III}^B = -1^\bullet$), ($Q_I^T = 0$, $Q_{II}^B = -1^\bullet$, $Q_{III}^T = 1^*$), and ($Q_I^B = 0$, $Q_{II}^B = 1^*$, $Q_{III}^T = 0$), respectively. Note that the notation for the topological charges is illustrated in the description of Table 2.

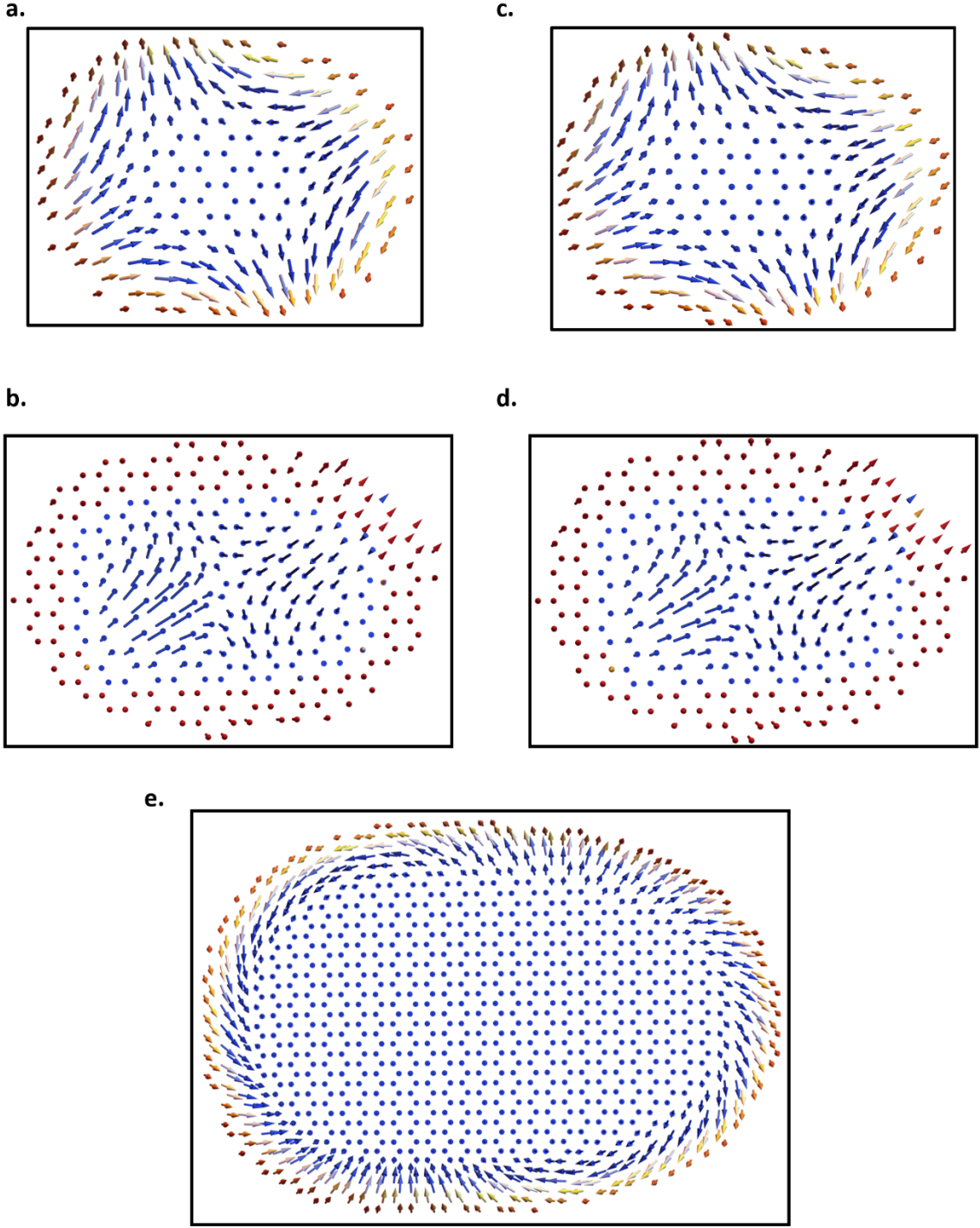

**Supplementary Figure 4. a, b** Field-assisted antiskyrmion (**a**) stabilized by an antivortex moiré interlayer field (**b**) in region II of the nonchiral Heisenberg-IDMI model with $\theta = 2.13°$ and $B = 0.5\,T$. Removing the magnetic field at $0K$ induces a slight relaxation of the spins (**c**) and the field (**d**). Generally, the TSTs reported in our study are stable without the need for a permanent magnetic field. **e** A magnetically stable bound state of two Bloch-type skyrmions with opposite helicities and topological charge $Q = -2$. The skyrmionic molecule emerged in region II of the nonchiral Heisenberg model with $\theta = 0.86°$ and $B = 0.5\,T$.

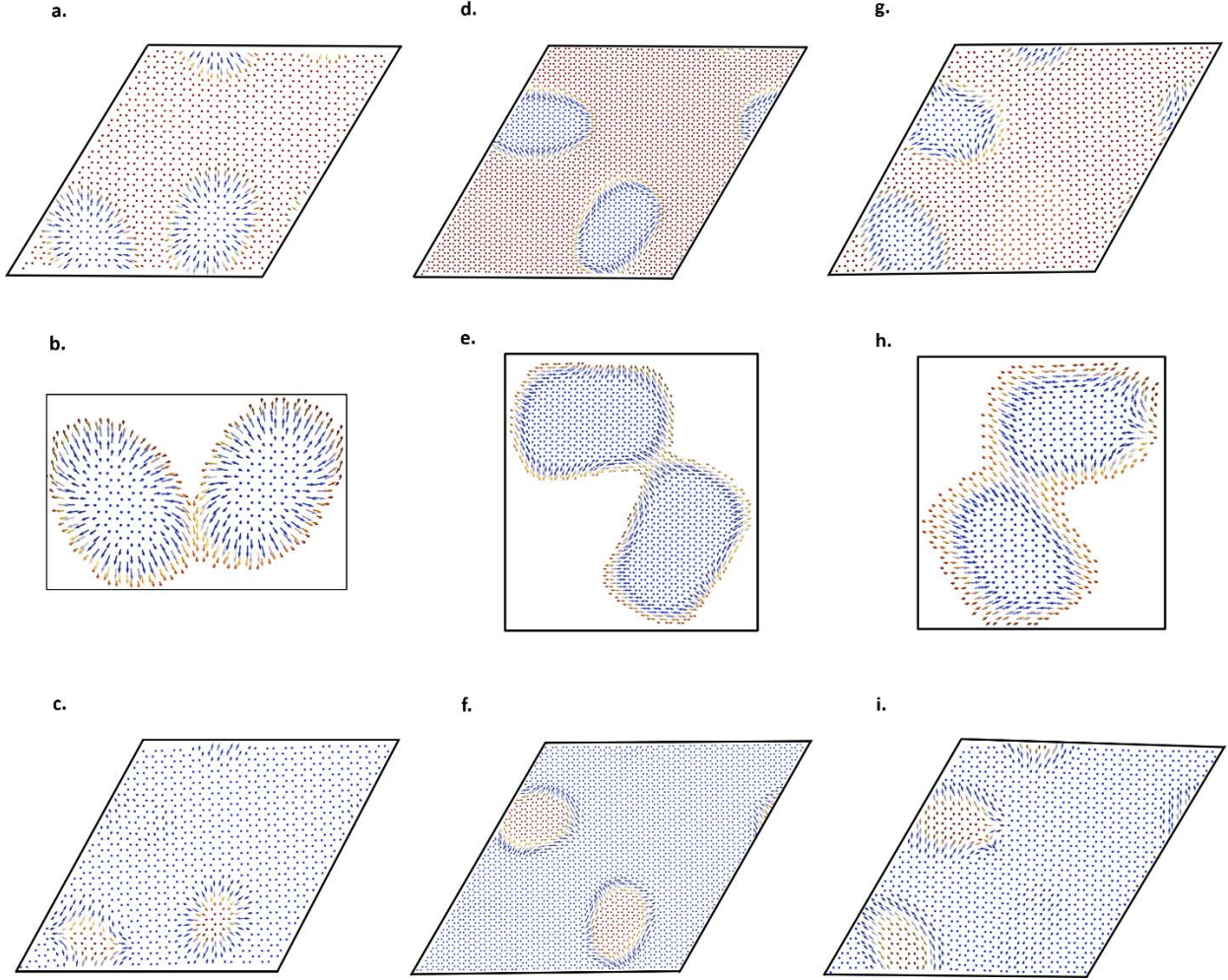

**Supplementary Figure 5. a, b, c** Tuning the ground state obtained in the chiral Heisenberg-IDMI model for $\theta = 2.45°$ and $B = 0.5T$. The initial state (**a**) has two Néel-type skyrmions trapped in regions I and III of the bottom layer. Applying an external magnetic field $\vec{B} = -2.2T\hat{z}$ at $0K$ couples the skyrmions (**b**). Increasing the magnetic field to $-2.7T\hat{z}$ reverses the magnetization (**c**) and stabilizes a final ground state with an antiskyrmion (region I) and a skyrmion (region III) in the top layer. The skyrmions have opposite helicities in the initial and final states. **d, e, f** Similarly, but for the nonchiral Heisenberg model with $\theta = 1.35°$ and $B = 0.75T$. The coupling (**e**) and reversal (**f**) are achieved at $-2.5T\hat{z}$ and $-3T\hat{z}$, respectively. The antiskyrmions in the initial (**d**) and reversed (**f**) ground states have opposite helicities. **g, h, i** The initial state (**g**) corresponds to the top layer of the nonchiral Heisenberg-IDMI model with $\theta = 2.13°$ and $B = 0.5T$. The coupling (**h**) and reversal (**i**) magnetic fields are $-1.7T\hat{z}$ and $-2T\hat{z}$, respectively.

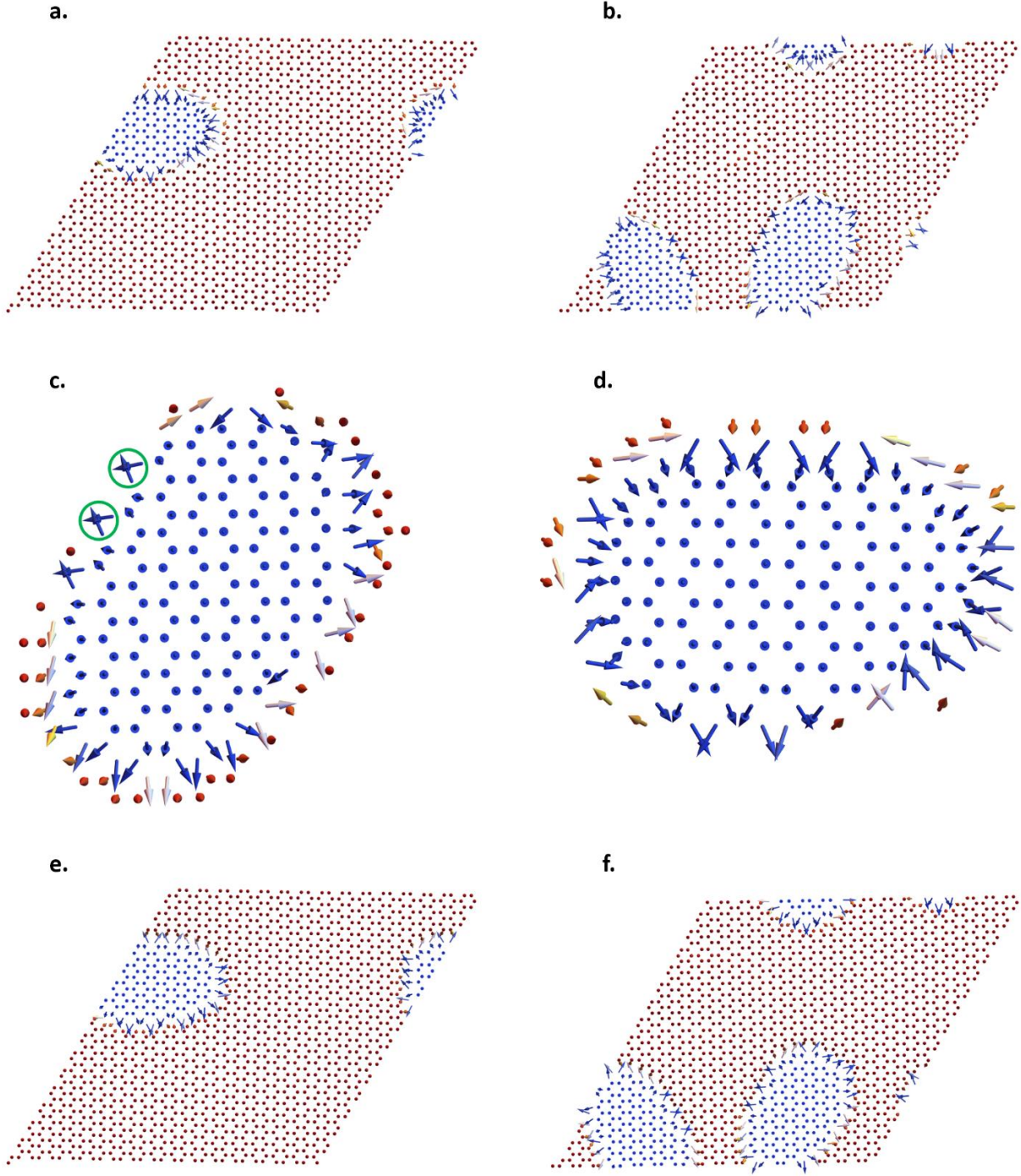

**Supplementary Figure 6. a, b** TSTs in the bottom (**a**) and top (**b**) layers for the nonchiral Heisenberg-Kitaev model with $\theta = 2.13°$ and $B = 0.5\, T$. The topological charges are $Q_I^B = 1$, $Q_{II}^T = -1$, $Q_{III}^B = 1$. **c, d** Illustration of the unconventional morphology for the TSTs in regions III and II, respectively. The green circles highlight the quasi-canting effect induced by the Kitaev interaction. The quasi-canting effect persists even in the presence of the chiral NN DMI (**e, f**).

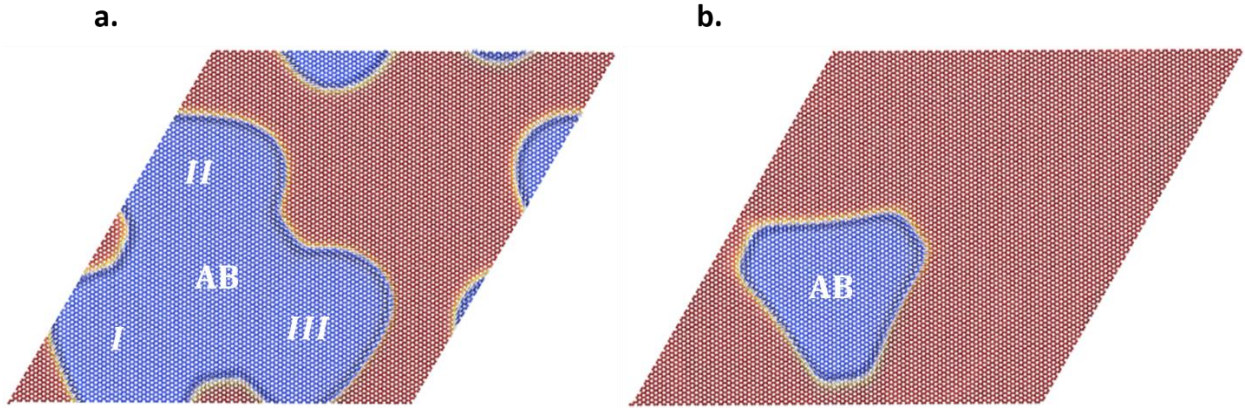

**Supplementary Figure 7. a** $Q = 1$ giant spontaneous TST with 9200 atoms (approximately) in the bottom layer of the nonchiral Heisenberg-IDMI model with $\theta = 0.65°$. In the top layer (**b**), a spontaneous TST with charge $Q = -1$ is trapped in the local AB stacking region of the moiré supercell with FM interlayer coupling.

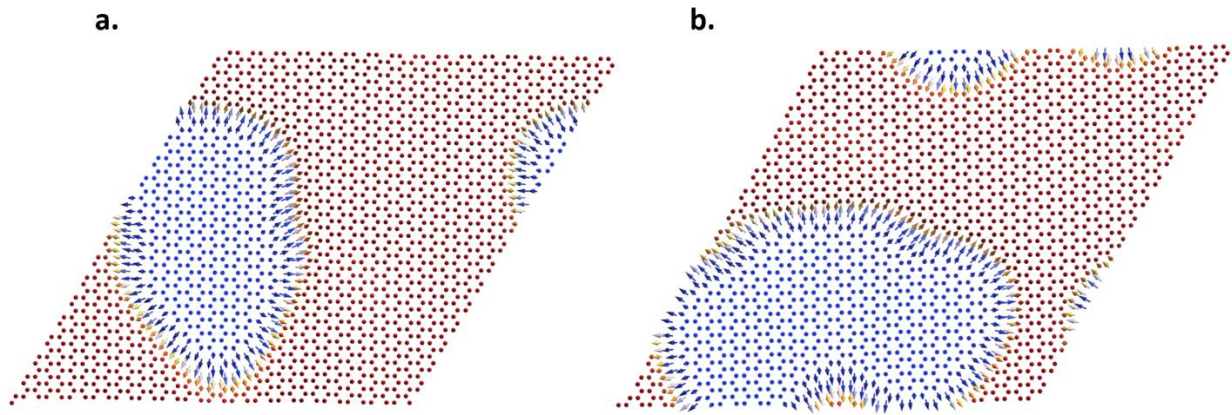

**Supplementary Figure 8.** The NN DMI locks the spontaneous merged TSTs to Néel-type skyrmions in the chiral Heisenberg and Heisenberg-IDMI models. The figures correspond to the bottom (**a**) and top (**b**) layers in the chiral Heisenberg model with $\theta = 1.89°$ and $B = 0\,T$.

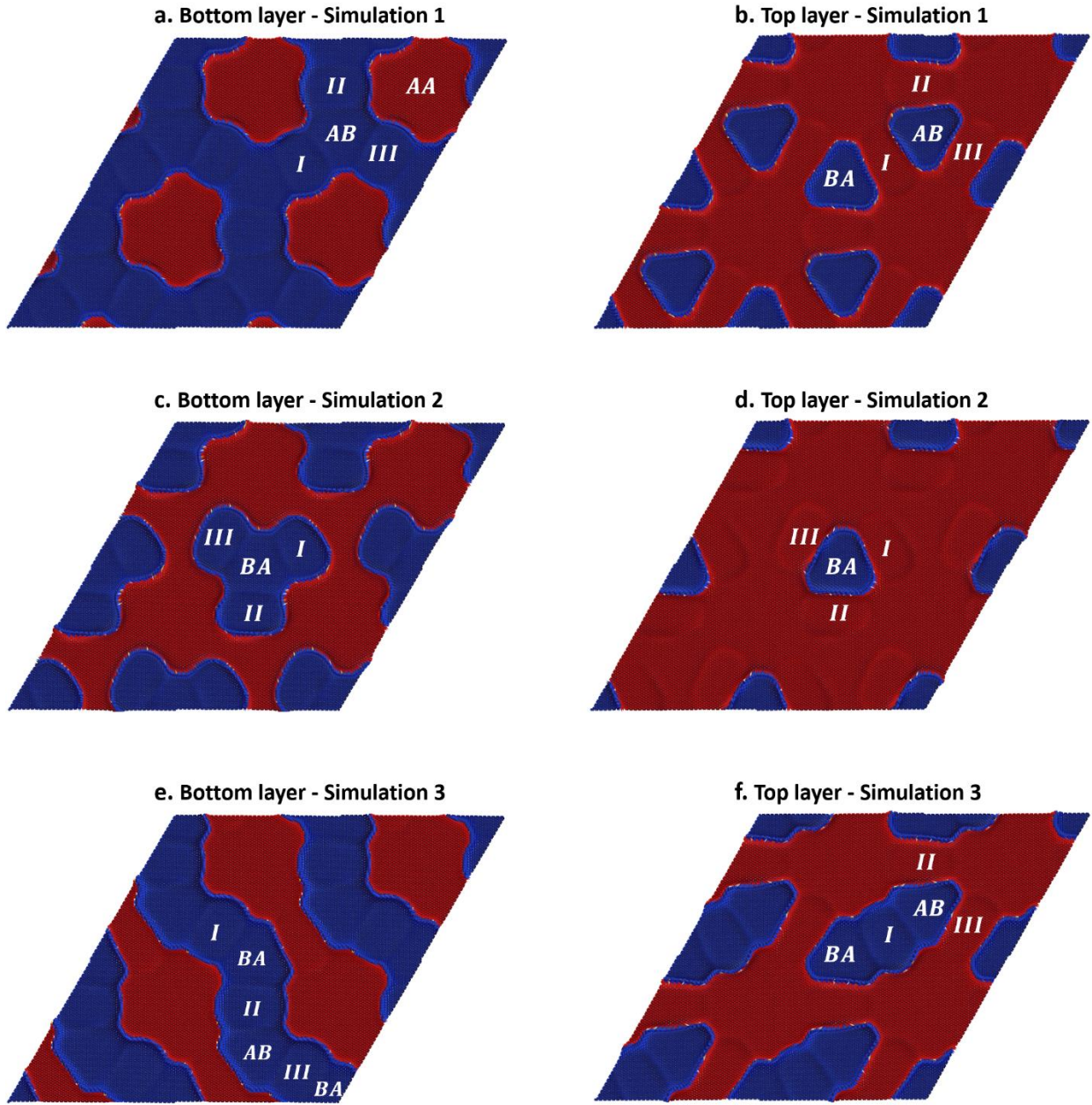

**Supplementary Figure 9.** We duplicate the TSTs reported in Figure 3 over a 2 × 2 moiré superlattice for illustration purposes (see caption in Figure 3). However, we stress that the moiré-periodicity of the magnetic ground state is broken in realistic multi-moiré supercell simulations. As discussed in the manuscript, adjacent moiré supercells have different initial states and display distinct ground state spontaneous spin textures.

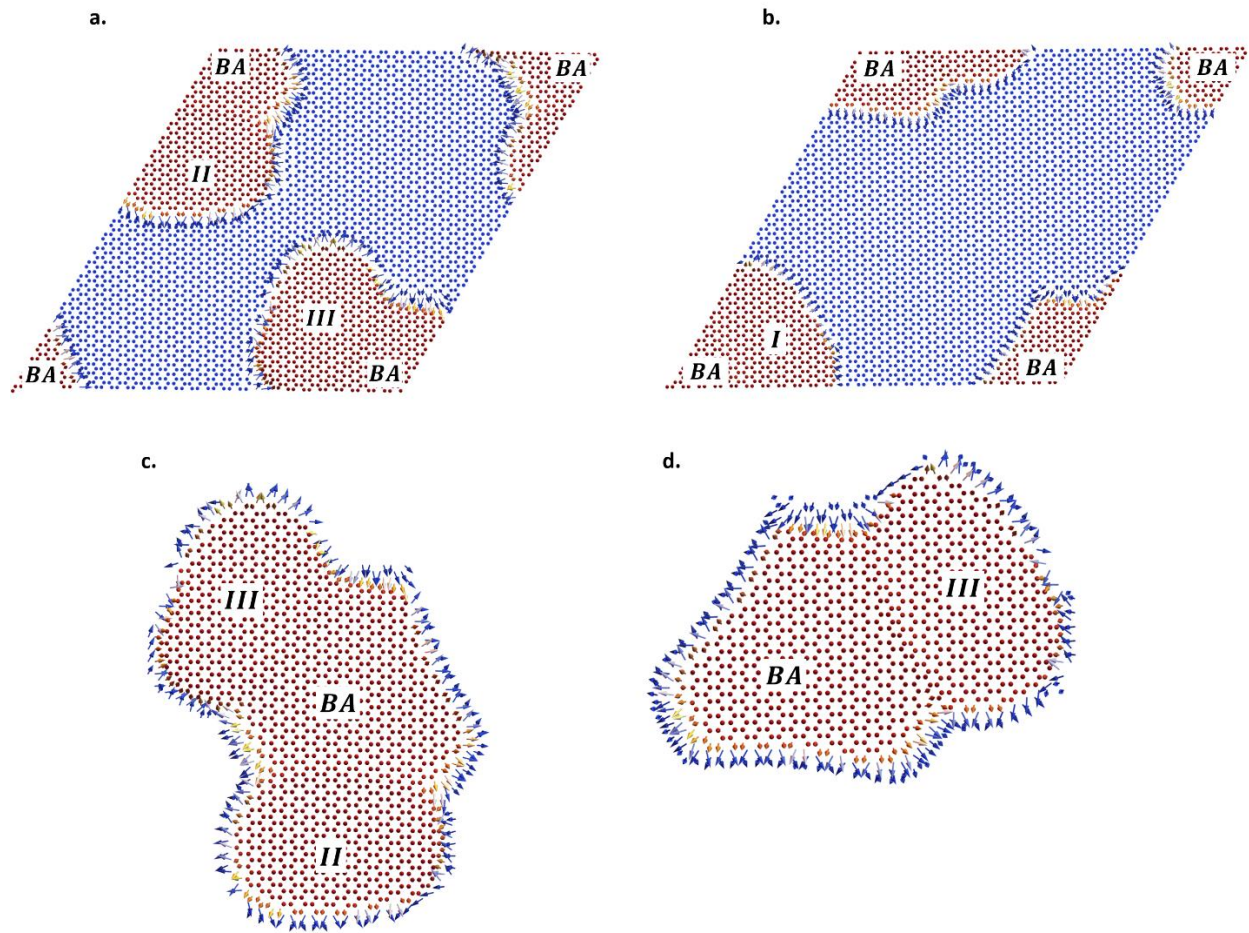

**Supplementary Figure 10. a** A spontaneous spin texture ($Q = 1$) in the bottom layer of the nonchiral Heisenberg-Kitaev model with $\theta = 0.76°$. The TST extends over the regions BA, II, and III. The Kitaev interaction induces a canting-like effect on the peripheral spins. **b** The top layer hosts a trivial magnetic bubble overlapping the regions III and BA. **c**, **d** Illustrate the unconventional morphologies of the TST and trivial MB, respectively.